\documentclass[pra,superscriptaddress,showpacs,twocolumn]{revtex4}

\usepackage{graphicx}
\usepackage{amsmath}
\usepackage{amssymb}
\usepackage{bm}
\usepackage{color}
\usepackage{maybemath}
\usepackage{longtable}

\usepackage{dcolumn}

\newcolumntype{j}{D{x}{}{17}}
\newcolumntype{h}{D{x}{}{22}}
\newcolumntype{g}{D{x}{}{27}}
\newcolumntype{f}{D{x}{}{33}}
\newcolumntype{.}{D{x}{}{-1}}

\renewcommand{\Re}{\mathrm{Re}\,}
\renewcommand{\Im}{\mathrm{Im}\,}

\newcommand{\dd}{\mathrm{d}}
\newcommand{\ee}{\mathrm{e}}
\newcommand{\ii}{\mathrm{i}}

\newcommand{\calO}{\mathcal{O}}
\newcommand{\PT}{$\mathcal{PT}\,$}

\newcommand{\ssc}{\mathrm{[sc]}}
\newcommand{\qqc}{\mathrm{[qc]}}
\begin{document}

\title{Imaginary Cubic Perturbation: Numerical and Analytic Study}

\author{Jean Zinn-Justin}
\email{jean.zinn-justin@cea.fr}
\affiliation{CEA/IRFU, Centre de Saclay, 91191 Gif-sur-Yvette Cedex, France}

\author{Ulrich D.~Jentschura}
\affiliation{Missouri University of Science and Technology,
Rolla, Missouri 65409-0640, USA}

\begin{abstract}
The analytic properties of the ground 
state resonance energy $E(g)$ of the cubic potential 
are investigated
as a function of the complex coupling parameter $g$.
We explicitly show that it is possible to 
analytically continue $E(g)$ by means of a resummed strong coupling 
expansion, to the second sheet of the Riemann surface, and we observe a 
merging of resonance and antiresonance eigenvalues at a critical point 
along the line $\arg(g) = 5 \pi/4$. In addition, we investigate the 
convergence of the resummed weak-coupling expansion in the strong coupling 
regime, by means of various modifications of order-dependent mappings 
(ODM), that take special properties of the cubic potential into account. 
The various ODM are adapted to different regimes of the coupling constant. 
We also determine a large number of terms of the strong coupling expansion by 
resumming the weak-coupling expansion using the ODM, demonstrating the 
interpolation between the two regimes made possible by this summation 
method.
\end{abstract}


\maketitle

%
% INTRODUCTION
%
\section{INTRODUCTION}
\label{intro}

The Hamiltonian
\begin{equation}
\label{eCubic}
H = - \frac12 \; \frac{\partial^2}{\partial x^2} 
+ \frac12 \; x^2 + \ii \; \frac{\sqrt{g}}{6} \; x^3 \,,
\end{equation}
with $g$ real positive, has been a subject of a number of studies since
it has been conjectured by Bessis and Zinn-Justin (1992) to have a real spectrum. Arguably, Eq.~\eqref{eCubic}
provides the simplest example of a \PT symmetric Hamiltonian, that is,
of a Hamiltonian invariant under a simultaneous 
complex conjugation (\textit{i.e.}, time reversal) and parity transformation
$x\mapsto -x$ (see Ref.~\cite{rBeBo}). In particular, the 
reality of its spectrum was proven first in the strong coupling limit \cite{rDDB},
then more generally 
in Ref.~\cite{rshin} (together with the positivity of the eigenvalues). 

Here, we continue our investigations~\cite{rJMPZJU} regarding its numerical
and analytic properties, including the analytic structure of 
its resonance energy eigenvalues as a function 
of the complex coupling parameter $g$, specially in the strong coupling regime and near the first level crossing singularity. 

We denote by $E(g)$ an energy eigenvalue of 
the Hamiltonian~\eqref{eCubic}.
One verifies that its perturbative expansion 
contains only integer powers of $g$:
\begin{equation}\label{eEnperturbative}
E(g) \mathop{\sim}_{g\to 0} \sum_{L=0} E_L \; g^L \,,
\end{equation}
with real coefficients $E_L$. The series is divergent for all~$g$, but 
is Borel summable~\cite{rCGM, rECali}, 
and a steepest descent calculation of the path integral
representation of the corresponding quantum partition function
\cite{rLOBLip, rLOBgen, rZJLOreport, rUJASJZJ} yields a large order behaviour
for the $n$th  energy eigenvalue of the form
\begin{equation}
\label{eLObehaviour}
E_{L} \mathop{\sim}_{L\to\infty}  (-1)^{L+1}\frac{6}{\pi^{3/2}}\frac{288^n}{n!} \;
  \;\frac{ \Gamma\left(L+ n+\tfrac12\right)}{A^{L+n+1/2}}\,,
\end{equation}
where $A=\tfrac{24}{5}$ is the instanton action.

More recently, Pad\'{e} summability was also rigorously established in
Ref.~\cite{rVGMMAM}.  The imaginary part of the energy levels on the
cut for $g=-|g|+\ii \; 0$ is positive. This proof confirms numerical
investigations based on the summation of the perturbative expansion by Pad\'{e}
approximants~\cite{rBeWe}.

More information on analytic and numerical properties was provided by an ODM
summation~\cite{rsezzin,rZJODM} of the perturbative series for the ground state
energy  (see Ref.~\cite{rJMPZJU}).  It was based on the mapping
$g\mapsto\lambda$,
\begin{equation}
\label{method3_1} 
g \; = \; \rho \; \frac{\lambda}{(1-\lambda)^{5/2}} \,,
\end{equation}
where $\rho$ is an order-dependent parameter.

In particular, the leading term of the strong coupling expansion could be
determined (and was found in agreement with results coming direct solutions of
Schr\"odinger equation \cite{rZnojil,rUJJZJxiii}) and the positivity of the
imaginary part verified numerically for all $g<0$. It was shown that along the
negative real axis, the imaginary part above the cut is a simple positive
decreasing function interpolating smoothly between a strong coupling power-law
behaviour and a non-analytic exponential tunneling factor obtained from
semi-classical instanton calculation for weak coupling.  Finally, it was argued
that the ground state energy is analytic up to $\arg (g) = 5\pi/4$ in the
second Riemann sheet, a property that we discuss in this article in more
detail.

By a simple coordinate translation, one can relate the Hamiltonian
\eqref{eCubic} to another \PT symmetric Hamiltonian, 
\begin{equation}
\label{eHamxiiixi}  
H^\qqc = - \frac12 \frac{\partial^2}{\partial x^2} 
+ \ii \; \left(\frac16 \; x^3 + \frac12 \, \chi \; x  \right) \,,
\end{equation}
where $\chi = g^{-4/5}$. This correspondence suggests to sum the 
series by another ODM of the form
\begin{equation}
\label{method1_1} 
g=\rho \; \frac{\lambda}{(1-\lambda)^{5/4}},
\end{equation}
which is shown here to be much more efficient in the strong coupling regime,
while the mapping~\eqref{method3_1} (and a variation
thereof discussed below) is more efficient for weak coupling.
The method~\eqref{method1_1} 
is used here in order to determine a number of coefficients of the large $g$
expansion (small $\chi$ expansion).
Finally, studying the limit of weak $\chi$, we also construct a
corresponding continued fraction (a subclass of Pad\'e approximants),
which enables us to obtain more insight 
into the analytic structure.

Combining all methods, we have a precise control  on the analytic continuation
of the ground state energy in the first Riemann sheet of the uniformization
variable $\chi = g^{-4/5}$, except near the cut in the $\chi$-plane.  Based on our
observation, we conjecture that the ground state energy  of the Hamiltonian
\eqref{eHamxiiixi} is a real analytic function, with a cut on the real negative
$\chi$ axis, which corresponds to a complex phase of $\arg(g) = 5 \pi / 4$ for
the original coupling parameter $g$ and thus lies on the second sheet of the
Riemann surface for the original coupling parameter $g$.  This domain is not
covered by Ref.~\cite{rshin}. The imaginary part of the
ground state energy on the cut for $\chi=-|\chi|+\ii \, 0$ is positive. The
small $\chi$ expansion of the resonance energies converges in a disk. From the
point of view of the initial Hamiltonian \eqref{eCubic}, the singularity
closest to the origin has the interpretation of a level crossing.

This can be
seen as follows: the cut for negative $\chi$ corresponds to a cut of the energy
level on the second sheet~\cite{rCBTTW,rEDDTT,rHKlWJ} 
of the Riemann surface (``Bender--Wu cut'').  It
starts from the limit of the circle of convergence of the strong coupling
expansion, i.e., from a point on the boundary of the circle of convergence of
the $\chi$ expansion. Two resonance energies are attached to points
infinitesimally displaced above and below the cut. The functions describing the
real parts of the two resonance energies are equal on the cut.
They cross as we cross the cut. By contrast, the functions
describing their imaginary parts are equal and opposite
on the cut and vary smoothly in its immediate vicinity.

At the start of the cut,
the imaginary part of the resonance and antiresonance energy is zero, but the
real parts still cross, and this configuration therefore corresponds to a
level crossing, given by the confluence of two energy levels on the second
sheet of the Riemann surface characterizing the resonance energies as a
function of $g$. For the Hamiltonian \eqref{eHamxiiixi}, it corresponds, in
addition, to a spontaneous breaking of the \PT symmetry (namely,
the spectrum composed of the resonance and
antiresonance energies with their equal and opposite
imaginary parts corresponds to a manifestly non-\PT symmetric
time evolution). Starting from the
region where the two eigenvalues merge and are real, we can say that after the
singularity, they become complex conjugate.

To confirm this picture, that is, analyticity in a cut-plane in the $\chi$
variable, sign of the imaginary part of the ground state energy on the cut and
level merging singularity we investigate together the ground state and the
first excited state, forming symmetric  combinations: sum and difference
squared of the corresponding energies.  We verify that these functions are
regular at the level merging singularity, showing that the level merging
involves the ground state and the first excited state. This gives us a totally
consistent picture and, moreover, allows us to determine with increased
precision a number of terms of the small $\chi$ expansion. In turn, from the
small $\chi$ expansion a continued fraction expansion is derived, whose
coefficients are all positive and seem to converge to a value consistent with a
square root singularity, within errors. 

As a final remark, we notice that the Hamiltonian \eqref{eCubic} can be
considered as a toy model for the $\ii\,\phi^3$ quantum field theory in two and
three space dimensions, whose renormalization group functions describe the
Lee--Yang edge singularity of Ising-like statistical models
\cite{rLYFisher,rBBJ}.

This paper is organized as follows.  We first discuss the scaling of the
coupling constant and the related strong coupling Hamiltonians in
Sec.~\ref{related}. General properties of ODM summation methods are recalled in
Sec.~\ref{summation}.  Optimized methods for the problem at hand are introduced
in Sec.~\ref{optimized}.  Numerical results for $g$ positive and negative are
obtained and discussed in Sec.~\ref{numerical}. Sec.~\ref{strongcoupling} is
devoted to the strong coupling expansion and level merging.  Analytic
properties of the ground state resonance in the second Riemann sheet and
continued fraction expansion are discussed in Sec.~\ref{analytic}.  Conclusions
are relegated to Sec.~\ref{conclu}.

%
% Related Hamiltonian: Strong coupling expansion
%
\section{STRONG COUPLING HAMILTONIANS}
\label{related}

The time-independent Schr\"{o}dinger equation
\begin{equation}
\label{eSchr}
\left( -\frac12 \frac{\partial^2}{\partial x^2} 
+ \frac12 \; x^2 
+ \frac{\ii}{6} \; \sqrt{g} \; x^3 
\right) \, \psi(x) = E(g) \; \psi(x)
\end{equation}
corresponds to the Hamiltonian~\eqref{eCubic}.
With proper boundary conditions at infinity, 
this equation determines the eigenvalues
$E(g)$  of the operator.

After rescaling the variable $x$ as 
$x \mapsto g^{-1/5}\, x$, the eigenvalue 
equation can be rewritten as follows
(the superscript [sc] stands for the
strong-coupling limit)

\begin{equation}
H^\ssc(\xi) \, \psi(x) = E^\ssc(\xi) \; \psi(x) \,,
\end{equation}
where 
\begin{equation}
\xi = g^{-2/5} \,,
\end{equation}
and the strong coupling Hamiltonian is
\begin{equation}
H^\ssc(\xi) = -\frac12 \frac{\partial^2}{\partial x^2} +
\frac{\ii}{6} \; x^3 + \frac12 \, \xi \; x^2 \,.
\end{equation}
The energies of the original and the strong coupling Hamiltonian
are related by
\begin{equation}
\label{exiiilargeg}
E(g) = g^{1/5} \; E^\ssc (g^{-2/5}) = g^{1/5} \; E^\ssc (\xi)\,.
\end{equation}
In particular, we have in leading order
\begin{equation}
E(g) \mathop{\sim}_{g \to \infty} g^{1/5} \; E^\ssc(0) \,,
\end{equation}
and the strong coupling expansion
\begin{equation}
\label{expssc}
E^\ssc(\xi) = \sum_{L=0}^\infty E^\ssc_L \; \xi^L \,.
\end{equation}
In the uniformizing variable $\sqrt{\xi} = g^{-1/5}$, 
the singularities of $E(g)$ in the complex plane correspond to level crossings,
whereas the cut of the original resonance energy,
defined as a function of $g$, is moved to $\xi = \infty$.

We then shift the coordinate $x \mapsto x+\ii \; \xi$. The equation becomes
\begin{equation}
\label{Eu}
\left( -\frac12 \frac{\partial^2}{\partial x^2} 
+ \frac{\ii}{6} x^3 +
\frac{\ii}{2} \xi^2 x - \frac{1}{3} \xi^3 \right) \psi(x)=
E^\ssc(\xi) \; \psi(x).
\end{equation}
We conclude that
\begin{equation}
\label{eODMxiiix} 
E^\ssc(\xi) = -\frac{1}{3} \, \xi^3 + E^\qqc(\xi^2) \, ,
\end{equation}
where $E^\qqc(\chi \equiv \xi^2)$ is a function of the 
 linear  coupling 
\begin{equation}
\chi = \xi^2 = g^{-4/5}\,,
\end{equation}
and thus an eigenvalue of the \PT symmetric 
Hamiltonian~\eqref{eHamxiiixi} 
corresponding to the Schr\"{o}dinger equation
\begin{equation}
\label{eODMxiiixi} 
H^\qqc(\chi)  \, \psi(x) = E^\qqc(\chi)  \, \psi(x) \,,
\end{equation}
where 
\begin{equation}
H^\qqc(\chi) = -\frac12 \frac{\partial^2}{\partial x^2} +
\ii \, \left( \frac{1}{6} \; x^3 +
\frac{1}{2} \, \chi \; x \right) \, .
\end{equation}
A generalization of the considerations leading to 
Eq.~\eqref{Eu} to the normalization of the cubic potential
used in Ref.~\cite{rJSLZ} is given in Appendix~\ref{num_evidence}.
This generalization is somewhat non-trivial because the 
precise form of the strong coupling Hamiltonian depends 
on the normalization of the initial perturbation. 
When translated to the conventions of Ref.~\cite{rJSLZ}, 
the coefficient of relative order $\xi^3$ becomes $1/108$
for all states of the spectrum; this
affords an explanation for the observation made originally in 
Table~2 of Ref.~\cite{rJSLZ}.

The expansion of the ground state
resonance energy $E^\qqc(\chi)$ in powers of $\chi$ reads
\begin{equation}
\label{exiiixismallv}
E^\qqc(\chi) = \sum_{L=0}^\infty E^\qqc_L \; \chi^L \,.
\end{equation}
Unlike the perturbative expansion in $g$, it
is convergent in a disk. Therefore,
$E(g)$ has a convergent large $g$ expansion of the form
\begin{align}
\label{exiiiglarge}
E(g) 
=& \; g^{1/5} \; \left( -\frac13 \, \xi^3 + E^\qqc(\chi) \right) 
\nonumber\\[2ex]
=& \; - \frac{1}{3g} + g^{1/5} \; 
\sum_{L = 0}^\infty E^\qqc_L \; g^{-4 L/5}.
\end{align}
These simple transformations show  that $E^\ssc(\xi)$ has only one term odd in
$\xi$, which moreover can be read off from 
Eq.~\eqref{exiiiglarge}. As a consequence, for $g$ large, the
relevant expansion variable is $g^{-4/5}$ rather than $g^{-2/5}$,
an observation that influences the construction of optimized
ODM summation methods for our problem. In  the
uniformizing variable $g^{-1/5}$, it is sufficient to determine the eigenvalues for
\begin{equation}
-\frac{\pi}{4} \le 
\arg \left( g^{-1/5} \right) \le 
\frac{\pi}{4} \,,
\end{equation}
in order to know them in the whole first Riemann sheet
of the $\chi$ variable ($-\pi \leq \arg(\chi) \leq \pi$).
Conversely, knowledge of the resonance eigenvalues for
$-\pi \leq \arg(\chi) \leq \pi$ implies a determination for
$-5\pi/4 \leq \arg(g) \leq 5\pi/4$ which exceeds the first Riemann sheet.

Indeed, the Hamiltonian \eqref{eODMxiiixi} is \PT symmetric for all real values
of the ``quadratic coupling'' $\chi$ but corresponds to the Hamiltonian
\eqref{eCubic} only in the case $\chi\ge 0$.  It is thus interesting to study
its spectrum also in the region $\chi<0$, which corresponds to the analytic
continuation $g=|g|\;\exp( 5 \, \ii \, \pi/4 )$.  Equation~\eqref{eODMxiiixi}
is also related to residues of poles of solutions of a linear system associated
to the first Painlev\'e equation \cite{rMaso}.

%
% Order-dependent mapping summation method
%
\section{GENERAL SUMMATION METHODS} 
\label{summation} 

%
% Idea of the method
%
\subsection{Idea of the method}

We here briefly recall the ODM summation method.  Details can be found
elsewhere~\cite{rsezzin,rZJODM}.  The method is based on some {\it a priori}
knowledge, or educated guess, of the analytic properties of the function that
is expanded. It applies both to convergent and divergent series, although it is
mainly useful in the latter situation. 

In what follows, we consider a function  analytic   in a sector   and
mappings   $g\mapsto \lambda$ of the form 
\begin{equation}
\label{eODMmapping} 
g=\rho \; \zeta(\lambda)   \,,
\end{equation}
where $\zeta(\lambda)$ is a real analytic function increasing on $0\le
\lambda<1$, such that $\zeta(\lambda) =\lambda+\calO(\lambda^2)$ and, for
$\lambda\to 1$, $\zeta(\lambda)\propto (1-\lambda)^{-\alpha}$ with $\alpha>1$.
A possible choice is 
\begin{equation}
g = \rho \; \frac{\lambda}{(1-\lambda)^\alpha} \, .
\end{equation}
The parameter $\alpha$  has to be chosen in accordance with the analytic
properties of the function $E$ (a rational number in our examples) and $\rho$
is an  adjustable parameter, whose interpretation is 
an order-dependent, ``artificial, local radius of convergence'' of the truncated,
divergent perturbative expansion (which becomes smaller
as the order of the transformation is increased).
An important property that singles out
relevant  mappings here is the following:  one chooses mappings such that,  for
$g\to\infty$ and thus $\lambda\to1$, 
the quantity  $g^{-1/\alpha}$ has a regular expansion
in powers of $1-\lambda$.  A more general discussion of the method can be found
in Ref.~\cite{rsezzin,rZJODM}. 

After the mapping~\eqref{eODMmapping},  $E$  is given by a Taylor series in
$\lambda$  of the form
\begin{equation}
E\bigl( g(\lambda)\bigr) =
\sum_{L=0}^\infty P_L(\rho) \; \lambda^L \,,
\end{equation}
where the coefficients $P_L(\rho)$  are polynomials of degree  $L$  in  $\rho$.
Since the result is formally independent of the parameter  $\rho$, the
parameter can be chosen freely. At  $\rho$ fixed, the series in $\lambda$ is
still divergent, but it has been verified on a number of examples (all Borel
summable), and proven in certain cases \cite{rRGKKHS} that, by adjusting
$\rho$ order by order, one can devise a convergent algorithm.  

The $K$th approximant  $E^{(K)}(g)$ is constructed in the following way: one
truncates the expansion at order  $K$  and chooses  $\rho$  as to cancel the
last term. Since  $P_K(\rho)$ has $K$ roots (real or complex), one chooses,
in general, for  $\rho$  the largest possible root (in modulus)  
$\rho = \rho_K$  with $P_K(\rho_K) = 0$, for which  $P_K'(\rho)$  is small. 
This leads to a sequence of approximants 
$E(g) \approx E^{(K)}(g)$, where 
\begin{equation}
E^{(K)}(g) = 
\sum_{L=0}^{K} P_L(\rho_K) \; \lambda^L(g,\rho_K) \,; \quad
P_K(\rho_K)=0 \, .
\end{equation}
The term $P_{K+1}(\rho_K)$ then gives an order of
magnitude of the error.

In the case of convergent series, it is expected that  $\rho_K$  has a
non-vanishing limit for $K\to\infty$.  By contrast, for divergent series it is
expected that $\rho_K$ goes to zero for large $K$ as 
\begin{equation}
\rho_K = \calO\left( E_K ^{-1/K}\right) \, . 
\end{equation}
The intuitive idea here is that $\rho_K$ corresponds to a
`local' radius of convergence. 

Let us conclude this section with a few remarks.  Above, we have used the
selection criterion $P_K(\rho_K)=0$ for the order-dependent $\rho_K$ which are
used to calculate the polynomials $P_L(\rho_K)$, for $L = 0, \ldots, K-1$.
Alternatively, one can choose the largest roots $\rho_K$ of the polynomials
$P'_K(\rho_K) = 0$  for which $P_K$ is small.  Other mixed criteria involving a
combination of $P_K$ and $P'_K$ can also be used. Indeed, one empirically
observes that the approximant is not very sensitive to the precise value of
$\rho_K$, within errors.  In the ODM method, the determination of the sequence of
the $\rho_K$'s is the most time-consuming computational step. Indeed, once the
$\rho_K$ are known, for each value of $g$, the calculation reduces to inverting
the mapping \eqref{eODMmapping} and simply summing the Taylor series in
$\lambda$ to the relevant order.

%
% Convergence analysis
%
\subsection{Convergence analysis}
\label{ssODMconvergence}

We here give a heuristic analysis~\cite{rsezzin,rJMPZJU} of the convergence of
the ODM method that shows how the convergence can be optimized. 
This will
justify the choice of the class of zeros of the polynomials $P_K$ or $P'_K$
and provide a quantitative analysis of the corresponding convergence.

In order to focus the analysis, we assume that we are dealing with functions
that have the properties of the eigenvalues of the Hamiltonian \eqref{eCubic}.
We consider only real functions analytic in a cut-plane with a cut along the
real negative axis and a Cauchy representation of the form
\begin{equation}\label{eECauchyrep}
E(g)=E(0)+  \frac{g}{\pi}\int^{0_-}_{-\infty}
\dd g' \; \frac{\Delta (g')}{g'(g'-g)} \,, 
\end{equation}
 where the subtraction  ensures the convergence of the integral for
$g \to -\infty$. 

For the  example  we discuss here, one can derive  by a steepest descent
calculation  \cite{rLOBLip,rLOBgen,rZJLOreport} an asymptotic behaviour of the
form
\begin{equation}
\label{eDgasymptotic}
\Delta (g) \mathop{\propto}_{g\to 0_-} 
\frac{1}{(-g)^{b+1}} \;
\exp\left( \frac{A}{g} \right)  \,,\qquad A>0\,.
\end{equation}
The function
$E(g)$ can be expanded in powers of $g$,
\begin{equation}
\label{edivergent1}
E(g) = \sum_{L=0}^\infty E_L \; g^L \,, \quad
E_L = \frac{1}{\pi} \,
\int_{-\infty}^{0_-} \frac{\dd g }{ g^{L+1}} \, \Delta (g)\,,
\end{equation}
the Cauchy representation of $E_L$ being valid for $L>0$. 
  
The asymptotic behaviour~\eqref{eDgasymptotic} then implies a large order
behaviour of the form~\eqref{eLObehaviour},
\begin{equation}
\label{edivergent2}
E_L \mathop{\propto}_{L\to\infty} 
(-A)^{-L} \; \Gamma(L+b+1) 
\mathop{\propto}_{k\to\infty} 
(-A)^{-L} \; L^{b} \; L!\,.
\end{equation}
A remark is in order.
When $g$ becomes negative in Eq.~\eqref{eSchr}, the expression 
$\ii \, \sqrt{g}$
becomes real, and the corresponding resonance and antiresonance eigenvalues are
attached to the upper and lower side of the integration contour which extends
from $g' = -\infty$ to $g' = 0$.  The discontinuity $\Delta(g')$ is the
difference of the imaginary parts of the resonance and antiresonance
eigenvalues. For the potential we study here, it has been proved that 
$\Delta(g')$ is positive \cite{rVGMMAM}, and this is consistent with
Eq.~\eqref{eDgasymptotic}.  In expression \eqref{eECauchyrep}, 
for $g>0$ the integrand is positive throughout the 
integration contour (the denominator contains a product of two
negative quantities).
The eigenvalues $E(g)$ of the \PT symmetric
Hamiltonian  \eqref{eCubic} thus are positive since  $E(0) = 1/2$ is positive.
Similarly, the expression~\eqref{edivergent1} shows that the sign of $E_L$ is
$(-1)^{L+1}$, a result consistent with the estimate 
given in Eq.~\eqref{edivergent2}.

\begin{figure}
\includegraphics[width=0.6\linewidth]{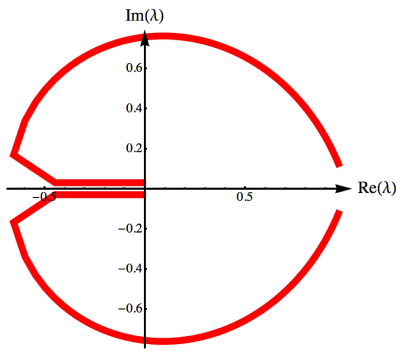}
\caption{\label{images1} (Color online.)
Image in the complex $\lambda$ plane
of the negative real $g$ axis under the
mapping $g = \rho \, \lambda/(1 - \lambda)^{5/2}$,
for $\rho = 1/2$. A point infinitesimally displaced
above the negative real axis is mapped onto a point
in the upper complex $\lambda$ plane.}
\end{figure}

\begin{figure}
\includegraphics[width=0.6\linewidth]{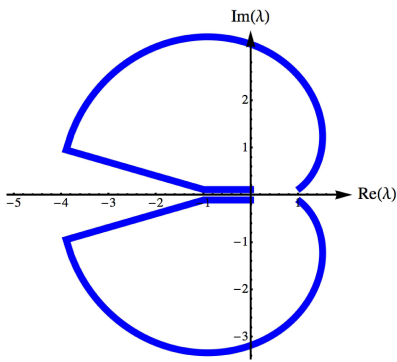}
\caption{\label{images2} (Color online.)
Same as Fig.~\ref{images1} but for the mapping
$g = \rho \, \lambda/(1 - \lambda)^{5/4}$.}
\end{figure}

We now introduce the order-dependent  mapping~\eqref{eODMmapping}, 
\begin{equation}
g = \rho \; \zeta(\lambda)\,,
\end{equation}
 with the understanding that the 
point $\lambda = 1$ corresponds to the point $g \to \infty$.
Because we can alternatively formulate the dispersion relation
in terms of $\lambda$,
the Cauchy representation then can be written as
\begin{equation}
E\bigl(g(\lambda )\bigr) = 
\frac{1}{\pi} \, \oint_{\Gamma} \dd \lambda' 
\frac{\Delta \bigl(g(\lambda ')\bigr)}{\lambda' -\lambda } \,,
\end{equation}
where $ \Gamma$ initially  is the image of the cut on the real negative axis
(see Fig.~\ref{images1} and~\ref{images2} for examples).
The contour $\Gamma$ can then be deformed if the function has analyticity
properties beyond the first Riemann sheet.

We expand
\begin{equation}
\label{eEODMlambda}
E\bigl(g(\lambda )\bigr) =
\sum_{K=0}^\infty P_K(\rho) \; [\lambda (g)]^K 
\end{equation}
with
\begin{equation}
\label{eODMPkgen} 
P_K(\rho) = \frac{1}{\pi} 
\oint_{\Gamma} \frac{\dd \lambda}{\lambda^{K+1}}  
\Delta \bigl(g(\lambda  )\bigr)\,.
\end{equation}
For $K \to \infty$, the factor $\lambda ^{-K}$ favours small 
values of $|\lambda|$,
but for too small values of $\lambda$, the exponential decay of 
$\Delta(g(\lambda))$ takes over.  Thus, $P_K(\rho_K)$ can be evaluated by the
steepest descent method. The ansatz  
\begin{equation}
\label{eODMansatz}  
\rho_K \mathop{\sim}_{K \to \infty} \; \frac{R}{K} \,,\quad R>0\,,
\end{equation}
implies that in the limit $K \to \infty$,
the saddle point values of $\lambda$ are independent of $K$ and
that $g(\lambda )\to0$. The former statement can easily been seen
by taking the logarithm of the integrand in
Eq.~\eqref{eODMPkgen} and observing that $K$ factors out in the
equation defining the saddle point.
Thus,  $\Delta (g)$ can be replaced by its asymptotic
form \eqref{eDgasymptotic} for $g\to0_-$, except for $g(\lambda)$
 of order one (or larger) 
and thus $\lambda$ close to the initial endpoint $\lambda =1$ for $K\to\infty$.
The contribution of the endpoint depends on the analytic properties of
$\Delta(g)$ but is bounded by a quantity of order
$\exp(C_1 K^{1-1/\alpha })$ [see Eq.~\eqref{elambdaK}].

In what follows we set
\begin{equation}
\frac{R}{A}=\mu\,,
\end{equation}
since $\mu$ is the only parameter (and it is independent of the normalization
of $g$). The behaviour of $P_K(\rho_K)$ is then  given by the sum of leading
saddle  contributions. Each saddle point contribution is of order
\begin{equation}
\label{eODMPkbehav}
\{P_{k}(\rho_K)\}_{\textrm{saddle\,point}} = \calO(\ee^{K\sigma })\,, \quad 
\sigma \equiv \frac{1}{\mu \; \zeta(\lambda)} - \ln(\lambda) \,,
\end{equation}
where $\lambda $ is determined by the saddle point equation 
\begin{equation}
\label{eODMsaddle} 
\sigma'(\lambda) =
- \frac{1}{\mu} \; 
\frac{\zeta'(\lambda)}{\zeta^2(\lambda)} -
\frac{1}{\lambda} =0\,.
\end{equation}
The analysis is simple only if $\lambda $ at the saddle point is real negative.
By contrast, if $\lambda $ is complex, the analytic properties of $\Delta (g)$
and  the possibility of deforming the contour $\Gamma $  become relevant.\par
For $\mu$ large, the equation  has a unique  solution ($\lambda\sim-1/\mu$),
which is real negative.  Then we note that at this saddle point, as a function
of $\mu$,
\begin{equation}
\frac{\partial \sigma}{\partial \mu} =
- \frac{1}{\mu^2 \; \zeta(\lambda)} > 0\,,
\end{equation}
as long as the saddle point value of $\lambda$ is negative. This suggests
decreasing $\mu$ as much as possible to improve the convergence.  The
exponential rate $\sigma$ corresponding to the saddle point vanishes for
\begin{equation}
\label{eODMsing} 
0 = \Re\bigl(\sigma(\lambda)\bigr) 
  = \frac{1}{\mu \; \zeta(\lambda)} -\ln(-\lambda) \,,
\end{equation}
and this defines a special value 
$\mu = \mu_c$ of the parameter $\mu$.  In
\cite{rsezzin,rZJODM,rJMPZJU}, we have discussed in some detail 
mappings of the form
\begin{equation}
\label{eODMalpha} 
g = \rho \; \frac{\lambda}{(1-\lambda)^{\alpha}},
\end{equation}
for various values of $\alpha$. 
Most notably, $\alpha=5/2$ has been applied
to the present problem (for a motivation regarding this mapping, 
see Appendix~\ref{paramodm}).

The system of Eqs.~\eqref{eODMsaddle} and \eqref{eODMsing}
can be solved for the variables $\mu = \mu_c$ and 
$\lambda = \lambda_c$. We obtain
\begin{subequations}
\begin{align}
\label{method1}
& \mbox{mapping (a):} \quad 
g = \; \rho \; \frac{\lambda}{(1-\lambda)^{5/4}} \,,
\\[1ex]
& \Rightarrow
\mu_c^{(a)} = 3.811\,522\ldots\,, \;\; 
\lambda_c^{(a)} = -0.259\,901\ldots\,,
\nonumber
\\[2ex]
\label{method2}
& \mbox{mapping (b):} \quad 
g = \rho \; \frac{\lambda \, (1-\lambda/2)}{(1-\lambda)^{5/2}} \,,
\\[1ex]
& \Rightarrow  
\mu_c^{(b)} = 4.445\,762\ldots \,, \;\; 
\lambda_c^{(b)} = -0.216\,262\ldots \,,
\nonumber
\\[2ex]
\label{method3}
& \mbox{mapping (c):} \quad 
g = \rho \; \frac{\lambda}{(1-\lambda)^{5/2}} \,,
\\[1ex]
& \Rightarrow  
\mu_c^{(c)} = 4.895\,690\ldots \,, \;\; 
\lambda_c^{(c)} = -0.189\,645\ldots \,.
\nonumber
\end{align}
\end{subequations}
Then, we expect the error $\varepsilon _K$ at order $K$ to be of 
the same order 
as the leading saddle point contribution, that is,
of order $P_K(\rho_K) \, \lambda(g,\rho_K) ^K$ or
in view of Eq.~\eqref{eODMPkbehav},
\begin{equation}
\varepsilon _K = \calO(\ee^{K\sigma} \; \lambda(g,\rho_K) ^K) \,.
\end{equation}
For $K\to\infty$, at $g$ fixed, $\lambda $ goes to 1 as
\begin{equation}
1-\lambda(g,\rho_K) \mathop{\sim}_{K\to \infty}
C_2(K g)^{-1/\alpha }\,,
\end{equation}
where for $\mu=\mu_c$ and methods \eqref{method1}, \eqref{method2} \eqref{method3}, 
respectively
\begin{subequations}
% \label{eODMCii}
\begin{align}
\label{C2_1}
C_2^{(a)} = & \; (A \, \mu_c^{(a)})^{4/5}   \approx 10.23\,, \\[2ex]
\label{C2_2}
C_2^{(b)} = & \; \left( \frac{A \, \mu_c^{(b)}}{2} \right)^{2/5} 
\approx 3.40\,, \\[2ex]
\label{C2_3}
C_2^{(c)} = & \; \left( \frac{A \, \mu_c^{(c)}}{2} \right)^{2/5} \approx 3.53\,.
\end{align}
\end{subequations}
Then,
\begin{equation}\label{elambdaK} 
\lambda ^K(g,\rho_K) \; 
\mathop{\sim}_{K\to\infty} \;
\exp\left( -C_2 K^{1-1/\alpha } g^{-1/\alpha } \right) \,.
\end{equation}
Then, two cases are possible:

Case (i): if the contribution to the integral corresponding to  $g(\lambda )$ of
order unity decreases exponentially with $K$ (which implies the possibility of
deforming the contour $\Gamma$), we can also choose $\mu<\mu_c$ and the ODM
method converges for all $g\ne 0$. Moreover, this in turn implies that the
function $E(g)$ is an entire function of $g^{-1/\alpha }$. This clearly is a
non-generic situation, but several examples have been met in the form of simple
integrals.
 
Case (ii): in a generic situation where the mapping removes the singularity at
infinity but other singularities are present, as explained in
Refs.~\cite{rsezzin,rZJODM,rJMPZJU} we then expect the optimal mapping to
correspond to $\mu=\mu_c$ and $P_K(\rho_K)$ (or $P_{K+1}(\rho_K)$ if we choose
for $\rho_K$ a zero of $P_K$) to be,  asymptotically for $K\to\infty$, itself
of order 
\begin{equation}
P_K(\rho_K) =
\calO\left[ \exp \left( C_3 \; K^{1-1/\alpha} \right) \right] 
\end{equation}
where $\alpha=5/4$ for method~\eqref{method1}
and $\alpha=5/2$ for methods~\eqref{method2} and~\eqref{method3}, respectively.

The domain of convergence then depends on the sign of the constant $C_3$.
For $C_3>0$, the domain of convergence is
\begin{equation}
|g| < \left( \frac{C_2}{C_3} \right)^{\alpha} \; 
\left[
\cos\left( \frac{{\rm arg}(g)}{\alpha} \right) \right]^\alpha \,. 
\end{equation}
For $\alpha=5/2$, this domain extends beyond the first Riemann sheet and
requires analyticity of the function $E(g)$ in the corresponding domain.

For $C_3<0$, the domain of convergence is the union of the sector 
$|\arg (g)| < \pi\alpha /2$ 
and the domain 
\begin{equation}
|g| > \left| \frac{C_2}{C_3} \right|^{\alpha} \; 
\left[ - \cos\left( \frac{{\rm arg}(g)}{\alpha} \right) 
\right]^\alpha \,. 
\end{equation}
Again for $\alpha =5/2$, this domain extends beyond the first
Riemann sheet. We will see that this is the situation realized in the three methods 
\eqref{method1},~\eqref{method2} and~\eqref{method3}.

One might now wonder how this analysis is related to the zeros of $P_K$ or
$P'_K$. Choosing the relevant zeros is a natural way of finding the region of
minimal values. In fact, the relevant zeros lie in the region where two leading
saddle points cancel or the leading saddle point(s) cancel the contribution
coming from the remaining part of the Cauchy integral.

Indeed, in the discussed examples of case (i), one could show that the zeros
correspond to a cancellation between saddle points. In the case (ii),
empirically one observes that the zeros are consistent with the asymptotic
behaviour \eqref{eODMansatz} but, in general, with power-law corrections: 
\begin{equation}
\rho_K= \frac{R}{K} \; \left(1+\delta_K\right),
\end{equation}
where $\delta _K\to0$ for $K\to\infty$.
A comparison with Eq.~\eqref{eODMPkbehav}
then leads to the following correction factor to the saddle point contribution 
\begin{equation}
\label{eODMmucorr}
\exp\left( -\frac{K \; \delta_K}{\mu_c \; \zeta(\lambda_c)} \right) =
\exp\bigl(K\delta_K\ln(-\lambda_c)\bigr) \, .
\end{equation}
Consistency with the preceding analysis, cancellation of the endpoint
contribution in the integral and convergence analysis, then requires
\begin{equation}
\delta_K=\calO(K^{-1/\alpha}) \,
\end{equation}
and this is also what is generally found.

%
% OPTIMIZED SUMMATION METHODS
%
\section{OPTIMIZED SUMMATION METHODS}
\label{optimized}

%
% ODM summation with $\alpha=5/4$
%
\subsection{ODM summation with $\maybebm{\alpha=5/4}$}
\label{optimizedmap}

We now consider the spectrum of the Hamiltonian \eqref{eCubic}, which can
be  obtained by solving the time-independent Schr\"{o}dinger equation
\begin{equation}
\label{eschrodingerxiii} 
\left( -\frac12 \frac{\partial^2}{\partial x^2} +
\frac12 x^2+
\frac{\ii}{6} \sqrt{g} \; x^3 \right)\psi(x) =
E \psi(x)
\end{equation}
with appropriate boundary conditions.

The spectrum has a perturbative expansion in integer powers of $g$ with
real (rational) coefficients alternating in sign, of the form
\eqref{eEnperturbative}, which can be derived up to large orders by standard
techniques.
Instanton calculus, based on a steepest descent evaluation of the corresponding
path integral, allows to 
calculate the classical action $A$ relevant for the large
order behaviour~\eqref{eLObehaviour}. One finds
\begin{equation}
\label{defA}
A=\frac{24}{5}\,.
\end{equation}

According to Eq.~\eqref{eODMxiiix}, the strong coupling expansion of an
eigenenergy $E(g)$ of the cubic contains a coefficient $g^{1/5} \, (-1/3) \,
g^{-6/5} = -\tfrac13 \, g^{-1}$. We can subtract this term and introduce the
function
\begin{equation}
\label{Fdef}
F(g)=\frac{1}{3}+g \; E(g) \,,
\end{equation}
which has a regular small $g$ expansion and a large $g$ expansion of the form
\begin{equation}
\label{eODMFxiii}
F(g) = g^{6/5} \; E^\qqc(g^{-4/5}) 
     = g^{6/5} \; \sum_{K=0}^\infty E^\qqc_K \; g^{-4K/5} \,,
\end{equation}
where the $E^\qqc_K$ coefficients have been
defined in Eq.~\eqref{exiiixismallv}.
Our aim is to investigate to which extent the strong-coupling 
expansion is described by an ODM summation of weak-coupling 
perturbation theory.
To this end, according to the discussion made in Appendix~A of 
Ref.~\cite{rJMPZJU}, we now introduce the mapping~\eqref{method1}
which reads
\begin{equation}
g=\rho \; \frac{\lambda}{(1-\lambda)^{5/4}} \, , 
\end{equation}
and set
\begin{equation}
F\bigl(g(\lambda)\bigr) = 
(1-\lambda)^{-3/2} \; \phi(\lambda,\rho).
\end{equation}
The function $\phi$ has a regular expansion both at $\lambda=0$ ($g=0$) and
$\lambda=1$ ($g\to\infty$).
The analysis of Sec.~\ref{summation} then 
yields $\mu_c^{(a)} =3.811\,522\ldots$ and thus
$R=A \, \mu_c^{(a)} = 18.295\,306\ldots$
in the notation of Sec.~\ref{ssODMconvergence}
[see Eq.~\eqref{method1}].

Although the summation methods apply to all eigenvalues, from now on we
consider only the \textit{ground state energy}. The first terms of
its perturbative expansion are
\begin{equation}
E(g) = \frac12 + \frac{11}{288} \; g 
  - \frac{930}{(288)^2} \; g^2+ \calO(g^3) \, .
\end{equation}
The convergence analysis of  Ref.~\cite{rJMPZJU}, 
partially reproduced in Sec.~\ref{summation}, clearly indicates that in this
problem, one expects a convergence of the ODM method 
only in $\ee^{-CK^\zeta}$ with $\zeta < 1$, and this is
confirmed here. We find strong evidence for a finite distance singularity.
It is consistent with this property that we can fit the zeros $\rho_K$ of
$P'_K(\rho)$ using the functional form
\begin{equation}
\label{rhok1}
\rho_K = \frac{A\mu_c^{(a)}}{K} \;
\left(1 - \frac{12.94}{(K+3)^{4/5} + 11.97} \right)\,.
\end{equation}
The zeros of $P_k$ can be fitted by an analogous formula but with slightly
different coefficients for the correction term.  It has proven convenient to
use the latter expression in the ODM summation in the place of the exact
values.

There is strong numerical evidence for a singularity at 
$g \approx 0.687\; \exp(5 \, \ii \, \pi/4)$.
(see Sec.~\ref{ssODMcontfrac}). One thus expects the 
ODM approximants of order $K$ to converge to the energy 
eigenvalues as (see Eq.~\eqref{C2_1})
\begin{equation}
\label{eODMviv}
\hbox{constant} \times 
\exp\left\{ \left[ -13.8-10.23 \; \Re(g^{-4/5}) \right] \; 
K^{1/5} \right\} \, ,
\end{equation}
consistent with a domain of convergence 
\begin{equation}
\label{domain1}
\Re\left(g^{-4/5}\right)=\Re\chi > -1.351 \,.
\end{equation}
Indeed, this expression reproduces well the region where the convergence factor
becomes close to unity.  The  domain of convergence contains a segment on the
real negative $\chi$ axis not contained in the convergence domain obtained by
the ODM method \eqref{method3} of Ref.~\cite{rJMPZJU}, which is
\begin{equation}
\Re\left(g^{-2/5}\right)=\Re\sqrt{\chi}>-0.0826\,. 
\end{equation}
The expression \eqref{eODMmucorr} combined with a fit of the relevant zeros of
the polynomials $P'_K$ yields a coefficient of about $17.4$ instead of $13.8$
in Eq.~\eqref{eODMviv}.   Finally, a rough 
numerical analysis of the
convergence toward $E^\qqc_0$ rather yields $\exp(-18 \, K^{1/5})$.  This
implies that either the asymptotic regime is reached very slowly, and this is
the most likely explanation, or our heuristic convergence analysis in this
particular example is  oversimplified. This problem requires a further
analysis. 

Compared with the ODM method of Ref.~\cite{rJMPZJU}, the convergence is expected
here to be slower at small $g$  and, eventually, at extremely large $K$.
At order 150, for $g$ real and positive,  the transition
occurs for $|g|$ slightly larger than 1. Of course, the method with
$\alpha=5/2$ also converges in a larger sector, up to $|{\rm arg}(g)| =5\pi/4$
instead of $|{\rm arg}(g)| =5\pi/8$ for $\alpha=5/4$.

Finally, because the convergence is smooth (unlike what happens with
$\alpha=5/2$ methods) the precision of the results can still be improved (see
Tables~\ref{table1}, \ref{table2}) by using repeatedly the algorithm that to
any sequence $S_n$ substitutes
\begin{equation}
\label{edelta}
S_n \mapsto \frac{S_nS_{n+2}-S_{n+1}^2}{S_n+S_{n+2}-2S_{n+1}} \,.
\end{equation}
Empirically, we find that it is most advantageous to
apply this method to the odd and even orders separately.  This is a variant of Aitken's
$\Delta^2$ process as it is known in the literature. Then at order 150, the
method \eqref{method1} remains the most efficient one for
large coupling $|g|\ge 1$.

% 
% ODM summation with $\maybebm{\alpha=5/2}$
% 
\subsection{ODM summation with $\maybebm{\alpha=5/2}$}
\label{ssODMnewmap}

In Ref.~\cite{rJMPZJU}, the ODM method with the mapping \eqref{method3},
\begin{equation} g = \frac{\lambda}{(1-\lambda)^{5/2}} \,, \end{equation} was
applied directly to the energy eigenvalue $E(g)$, which has a large-order
expansion of the form
\begin{equation}
\label{eODME}
E(g) = g^{1/5} \; \sum_{K=0}^\infty E^\ssc_K \; g^{-2k/5} \,.
\end{equation}
This expansion effectively becomes a large-order expansion in powers of
$g^{-4/5}$ if one subtracts the term of order $-\tfrac13 \, g^{-6/5}$.  The
latter is the only term not accounted for by the powers of $g^{-4/5}$. 

The numerical results in Sec.~\ref{numerical} justify the introduction of the
function $F(g)$, which is defined in Eq.~\eqref{eODMFxiii}, and the mapping
\eqref{method1} that precisely takes the special form of the large $g$
expansion into account.  However, while the method~\eqref{method1} is more
efficient for $g$ large, the method~\eqref{method3} yields more precise results
for $K$ very large and converges in a larger sector of the complex plane. 

It is thus natural to investigate an ODM method,
still with $\alpha=5/2$, that
incorporates in a better way the properties of the large $g$ expansion than the
method given by Eq.~\eqref{method3}. 
To this end, we again introduce the function
  \eqref{Fdef}  and modify the form of the mapping to transfer the property
of even powers of $g^{-2/5}$ into even powers of $1 - \lambda$. A possible
choice for a modified mapping is the mapping \eqref{method2}, 
\begin{align}
g =& \; \rho \;
\frac{\lambda(1-\lambda/2)}{(1-\lambda)^{5/2}}
\\[2ex]
\Rightarrow \quad 
\left( \frac{2g}{\rho} \right)^{-2/5} =& \; (1-\lambda) \;
\left[1- (1-\lambda)^2 \right]^{-2/5} \, .
\nonumber
\end{align}
Also,
\begin{equation}
(1-\lambda)^{-3} =
\left( \frac{2g}{\rho} \right)^{6/5} \;
\left[1-(1-\lambda)^2\right]^{-6/5}.
\end{equation}
With this mapping, the function 
\begin{equation}
\phi(\lambda) = (1-\lambda)^{3} \; F\bigl(g(\lambda)\bigr)  
\end{equation}
is indeed only a function of
$(1-\lambda)^2$. However, in contrast to the previous mapping, the symmetry
$g^{-2/5}\mapsto -g^{-2/5}$ is only implemented approximatively, at a finite
order,  because the series in powers of $\lambda$ is truncated.

Since one cannot expect geometric convergence of the ODM method defined
by~Eq.~\eqref{method2}, the relevant value of $\mu$ is
$\mu_c^{(b)}=4.445\,762\,607\ldots$ (see Sec.~\ref{ssODMconvergence}).  It follows
that $C^{(b)}_2=(R/2)^{2/5}= (A \, \mu_c^{(b)}/2)^{2/5} = 3.401\,535\ldots$ [see
Eq.~\eqref{C2_2}].

For the ground state energy, we empirically find that the 
location of the zeros of $P'_K$ can be fitted by 
\begin{equation}
\label{rhok2}
\rho_K = \frac{A \; \mu_c}{K} \;
\left(1- \frac{5.0}{(K + 2)^{4/5}+4.6} \right).
\end{equation}
One also expects a term decreasing only like $K^{-2/5}$ but its coefficient is
apparently too small to be detected compared to the larger $K^{-4/5}$
contribution. The same remark actually applies to the method~\eqref{method3}.
We have not compared the convergence of methods~\eqref{method2}
to method~\eqref{method3} systematically, but
methods~\eqref{method2}
is definitively better  for $g\to\infty$ than with the initial $\alpha=5/2$
mapping of Eq.~\eqref{method3} 
(see Tables \ref{table1} and~\ref{table2}). For
example, at order $55$, with the new 
mapping~\eqref{method2}, 
one finds the value $E_{0}^\qqc=0.372\,545\,(9)$
for the leading strong-coupling coefficient
defined according to Eq.~\eqref{exiiixismallv}, corresponding to
a  relative error of the order of  $10^{-7}$, that is, two order of magnitudes
better as compared to the mapping~\eqref{method3}.
At order 150, one finally obtains $E_{0}^\qqc=0.372\,545\,78(6)$, a value that
differs by $4 \times 10^{-9}$ from the strong coupling limit in \eqref{eODMvexpand}.

Values of the ground state energy $E(g)$ for various real, positive and
negative values of $g$, and order 55 and 150, are reported in
Tables~\ref{table1}---\ref{table4}.

The general conclusion is: at the order we calculate, as expected, 
the results for the modified method~\eqref{method2} are
more precise than for the more straightforward method~\eqref{method3} if 
$g$ is large, but comparable or worse for $g$ small. 
However, for   $|g|\ge 1$, results given by the
method~\eqref{method1} with $\alpha=5/4$ are always the best.
This is consistent with the fact that
higher-order corrections of the form $g^{-4 K/5}$ are well
represented by the corresponding expansion for 
$E^\qqc(g)$. Note that the correction of relative order
$g^{-6/5}$ for $E^\ssc(g)$ provides for a certain irregularity 
in the expansion~\eqref{expssc}. As a conclusion, the advantages of
the modified version~\eqref{method2} 
seem to be somewhat limited, and we have not investigated
the method more thoroughly, but we have included the results in the tables to stress the general consistency of the different implementations of the ODM summation methods. 

%
% ODM summation: Numerical results
%
\section{NUMERICAL RESULTS: GROUND STATE ENERGY}
\label{numerical}

%
% Strong coupling expansion
%
We first
display a few typical results obtained for $g$ finite, positive and negative,
by the various methods we discuss in this article:  the ODM method with
$\alpha=5/4$,  the methods of Sec.~\ref{ssODMnewmap} and
Ref.~\cite{rJMPZJU} and, finally, the continued fraction of
Sec.~\ref{ssODMcontfrac} below. 

\begin{table*}[htb]
\begin{center}
\begin{minipage}{17.8cm}
\begin{center}
\caption{\label{table1} The ODM mappings given in
Eq.~\eqref{method1},~\eqref{method2} and~\eqref{method3} are compared with
respect to their convergence properties on the real positive axis at a
transformation order of $55$. Method~\eqref{method1} with $\alpha = 5/4$ and
extrapolation clearly provides the best answers for large and moderate
coupling.  The modified method~\eqref{method2} with $\alpha = 5/2$ interpolates
between method~\eqref{method1} and~\eqref{method3}; indeed, the application of
the straightforward ODM method~\eqref{method3} still provides for the best
convergence at weak coupling.} 
\begin{tabular}{c@{\hspace*{0.3cm}}fg}
\hline
\hline
\rule[-2mm]{0mm}{6mm}
$g$ & $0.5$ & $1$ \\
\hline
\rule[-2mm]{0mm}{6mm}
Method of Eq.~\eqref{method1} &
  0.51689\,x17642\,53171\,97821\,11588\,95(6)  &
  0.53078\,x17593\,04176\,67113\,556(1) \\
\rule[-2mm]{0mm}{6mm}
Method of Eq.~\eqref{method2} &
  0.51689\,x17642\,53171\,97821\,11588(9) &
  0.53078\,x17593\,04176\,67113(4) \\ 
\rule[-2mm]{0mm}{6mm}
Method of Eq.~\eqref{method3} &
  0.51689\,x17642\,53171\,97821\,11588\,9(6) &
  0.53078\,x17593\,04176\,671135(6) \\
\hline
\hline
\rule[-2mm]{0mm}{6mm}
$g$ & $5.0$ & $21.6$ \\
\hline
\rule[-2mm]{0mm}{6mm}
Method of Eq.~\eqref{method1} &
  0.60168\,x39332\,05191\,96159(1) &
  0.73340\,x99204\,85427\,96(4) \\
\rule[-2mm]{0mm}{6mm}
Method of Eq.~\eqref{method2} &
  0.60168\,x39332\,05(2) &
  0.73340\,x9920(5) \\
\rule[-2mm]{0mm}{6mm}
Method of Eq.~\eqref{method3} &
  0.60168\,x39332\,05(2) &
  0.73340\,x992(1) \\
\hline
\hline
\end{tabular}
\end{center}
\end{minipage}
\end{center}
\end{table*}
\begin{table*} 
\begin{center}
\begin{minipage}{16cm}
\begin{center}
\caption{\label{table2} Same as Table~\ref{table1},
but for a transformation order of~150. For weak and moderate coupling, the method \eqref{method3} now takes over. For 
strong coupling, the convergence of method~\eqref{method1} 
with $\alpha = 5/4$ clearly is superior to that of the 
other ODM methods. For completeness, we give also the results obtained from the simple continued fraction method \eqref{econtfracglarge}. 
For the method defined by~\eqref{method1}, the numerical
results are post-accelerated using the algorithm~\eqref{edelta}.}
\begin{tabular}{c@{\hspace{0.5cm}}gg}
\hline
\hline
\rule[-2mm]{0mm}{6mm}
$g$ & \multicolumn{1}{c}{0.5} & \multicolumn{1}{c}{1.0}  \\
\hline
\rule[-2mm]{0mm}{6mm}
Method of Eq.~\eqref{method1} &
  0.51689\,x17642\,53171\,97821\,11588\,9 & 
  0.53078\,x17593\,04176\,67113\,55618\,18032 \\
\rule[-2mm]{0mm}{6mm}
Method of Eq.~\eqref{method2} &
  0.51689\,x17642\,53171\,97821\,11588\,9 &
  0.53078\,x17593\,04176\,67113\,55618\,18032 \\
\rule[-2mm]{0mm}{6mm}
Method of Eq.~\eqref{method3} &
  0.51689\,x17642\,53171\,97821\,11588\,9 &
  0.53078\,x17593\,04176\,67113\,55618\,18032 \\
\rule[-2mm]{0mm}{6mm}
Continued fraction &
  0.51689\,x17642\,53171\,97821\,1(0) &
  0.53078\,x17593\,04176\,67113\,55(7) \\
\hline
\hline
\rule[-2mm]{0mm}{6mm}
$g$ & \multicolumn{1}{c}{5.0} & \multicolumn{1}{c}{21.6}  \\
\hline
\rule[-2mm]{0mm}{6mm}
Method of Eq.~\eqref{method1} &
  0.60168\,x39332\,05191\,96158\,93564\,94(4) &
  0.73340\,x99204\,85427\,96459\,24020(0) \\
\rule[-2mm]{0mm}{6mm}
Method of Eq.~\eqref{method2} &
  0.60168\,x39332\,05191\,96159(0) & 
  0.73340\,x99204\,8542(8) \\
\rule[-2mm]{0mm}{6mm}
Method of Eq.~\eqref{method3} &
  0.60168\,x39332\,05191\,96158(9) &
  0.73340\,x99204\,854(3) \\
\rule[-2mm]{0mm}{6mm}
Continued fraction &
  0.60168\,x39332\,05191\,96158\,936(0) &
  0.73340\,x99204\,85427\,96459\,240(3) \\
\hline
\hline
\end{tabular}
\end{center}
\end{minipage}
\end{center}
\end{table*} 

% 
% Real positive axis
% 
\subsection{Real positive axis}

For $g$ finite and real positive, we have chosen the same values as in
Ref.~\cite{rJMPZJU} and added results for the case  $g=0.5$. Most results are
displayed in Table~\ref{table1} for the results at order 55, to allow for a
direct comparison with \cite{rJMPZJU} and give an idea about the rate of
convergence, and in Table~\ref{table2} at order 150. We use everywhere the
convention  that the numerical uncertainty applies to the digit in parentheses
(this may also affect the preceding digit by a shift $\pm1$).

For $g=0.5$, at order 150, with the ODM method 
given in Eq.~\eqref{method2}, one finds
\begin{align}
E(0.5) = 0.&51689\,17642\,53171\,97821\,11588
\nonumber\\
& 95662\,1775(8) \,,
\end{align}
and with the method given in Eq.~\eqref{method3}
\begin{align}
E(0.5) =0.& 51689\,17642\,53171\,97821\,11588
\nonumber\\
& 95662\,17760\,99999\,61207\,(4) \,,
\end{align}
the error being in agreement with the estimate given in 
Ref.~\cite{rJMPZJU}.
Finally, with the method given Eq.~\eqref{method1} one finds
\begin{align}
E(0.5)=0.&51689\,17642\,53171\,97821\,11588\nonumber\\&95662\,1775(8)
\end{align}
and after extrapolation using the method~\eqref{edelta}, 
\begin{align}
E(0.5)=0.&51689\,17642\,53171\,97821\,11588\nonumber\\
& 95662\,17760\,99999\,612(1)\,.
\end{align}
For $g=1$, from a numerical solution of the 
Schr\"{o}dinger equation one obtains
\begin{align}
\label{E11}
E(1.0)=0.& 53078\,17593\,04176\,67113
\nonumber\\
& 55618\,18032\,225\,.
\end{align}
At order 150, with the mapping~\eqref{method3}, the 
result  reads
\begin{align}
\label{E12}
E(1.0)=0.&53078\,17593\,04176\,67113
\nonumber\\
& 55618\,18032\,22595(1) \, .
\end{align}
At order 150, the method~\eqref{method2} yields a result with an error of
$10^{-33}$, which is slightly less precise.
The method~\eqref{method1} together with extrapolation \eqref{edelta}
yields
\begin{align}
E(1.0)=0.&53078\,17593\,04176\,67113\,55618\,18032\nonumber\\&22595\,1(1) .
\end{align}
Clearly, for small values of $g$, that is, $g < 1$, at order 150 the method
of~\cite{rJMPZJU} gives the most precise results.  The entries for $E(1.0)$
exceed the maximum number of columns available in Table~\ref{table2}; for
results obtained at transformation order 55, see Table~\ref{table1}.

Finally, to compare with the Pad\'e summation of \cite{rBeWe}, we give  results for
$g=288/49$, which corresponds to $\lambda^2 = 1/7$ in 
\cite{rBeWe}.  In Ref.~\cite{rBeWe}, at
order 192, the reported result is $E =0.61273\,81063\,88986$. 
In Ref.~\cite{rJMPZJU}, using the 
method~\eqref{method3}, at order 55, one finds
\begin{equation}
E(\tfrac{288}{49}) = 0.61273\,81063\,89(1) \,. 
\end{equation}
With $\alpha=5/4$, using method~\eqref{method1} and convergence acceleration, at order 55, one obtains
\begin{equation}
E(\tfrac{288}{49}) = 
0.61273\,81063\,88984\,124(7) \,.
\end{equation}
This improves over the ODM method \eqref{method3}  in 
Ref.~\cite{rJMPZJU} by six
orders of magnitude. Then, at order~150, the result becomes
\begin{equation}
E(\tfrac{288}{49}) = 
0.61273\,81063\,88984\,12476\,20895\,52(6) \,.
\end{equation}
With the method~\eqref{method3}, at the same order~150, one obtains
\begin{equation}
E(\tfrac{288}{49}) = 
0.61273\,81063\,88984\,12476(3) \,.
\end{equation}
We have also verified that 
\begin{equation}
E(\tfrac{288}{49}) = 
0.61273\,81063\,88984\,12476\dots
\end{equation}
by a numerical solution of the Schr\"{o}dinger equation.

The general conclusion is: for values $g \le 1$, at order 150 the method  with
the mapping~\eqref{method3} of Ref.~\cite{rJMPZJU} is the most precise, and the
advantage increases with increasing order and decreasing parameter $g$. By
contrast, for $g>1$, at least up to order 150, the method with $\alpha=5/4$
takes over.  Note that from now on and, in particular in the tables, we only
quote   the results obtained by the method \eqref{method1} after extrapolation.

Finally, the continued fraction of Sec.~\ref{ssODMcontfrac}, constructed from
the strong coupling expansion (but which incorporates additional information
about level merging), provides a rather good representation of the function in
a wide domain.

For completeness, in the spirit of reference  
\cite{rBeWe}, we give some indication about
the summation of the perturbative series by a continued fraction expansion.
We define (for the ground state)
\begin{equation}
h_0(g)=E(g)/E(0),
\end{equation}
and introduce the relation 
\begin{equation}
h_{p-1}(g) = 1 + \frac{ \kappa _p \, g}{h_p(g)} \,, \quad h_p(0)=1 \,,
\end{equation}
which allows calculating $h_p$ from $h_{p-1}$ 
and determining recursively the coefficients $\kappa_p$.
The truncated continued
fraction (obtained by replacing, at some order $p$, $h_p(g)$ by 1) generates, 
alternatively, $[n+1/n]$ and $[n/n]$ Pad\'e approximants.
For a Stieltjes function, all coefficients $ \kappa _p$ are positive. Moreover, for a
divergent series with a large order behaviour of the form \eqref{eLObehaviour}
one expects the coefficients $ \kappa _p$ to grow asymptotically linearly with $p$ and
the error for the continued fraction truncated at order $p$ to behave like
$\exp(-C\sqrt{p/g})$. This is indeed what is observed. The coefficient of the
linear term is compatible with $5/48$. More precisely, a good fit is 
$\kappa _p =(10\,p+3\times(-1)^p)/96+ \calO(1/\sqrt{p})$ for $p\to\infty$.
Also, the coefficient of $\sqrt{p/g}$ extrapolates to
$C=2 \sqrt{48/5}=6.19\ldots$ with a good precision. Compared to the various ODM
summations, the convergence is limited to the first Riemann sheet, is poorer
than the ODM method with $\alpha=5/2$ for $g$ small and is much poorer than the
method with $\alpha=5/4$ for $g$ large. For example, at order $150$, for
$g=0.5$ the error is about $1.5\times 10^{-42}$, for $g=1$ about $6 \times 10^{-31}$,
for $g=5$ about $10^{-14}$ and about $2 \times 10^{-7}$ for $g=21.6$.

\begin{table*}
\begin{center}
\begin{minipage}{17cm}
\begin{center}
\caption{\label{table3}
On the real negative axis, at order~150,
the real part $\Re\, E(g)$ is approximated by several
variants of ODM methods. The    straightforward ODM 
method~\eqref{method3} provides the best results
for weak coupling, whereas method~\eqref{method1}
with $\alpha = 5/4$,   post-accelerated, is superior in the moderate and strong coupling domain. Note that the simple continued fraction
method \eqref{econtfracglarge} also provides for a  well converged answer.} 
\begin{tabular}{c@{\hspace{0.5cm}}gg}%{c..}
\hline
\hline
\rule[-2mm]{0mm}{6mm}
$-g$ & 
\multicolumn{1}{c}{$0.5$} &
\multicolumn{1}{c}{$1.0$} \\
\hline
\rule[-2mm]{0mm}{6mm}
Method of Eq.~\eqref{method1} &
  0.x47642\,7408(3)& 
  0.x44252\,00451\,24688(4) \\
\rule[-2mm]{0mm}{6mm}
Method of Eq.~\eqref{method2} &
  0.x47642\,74083\,2(7) & 
  0.x44252\,00451\,2(4) \\
\rule[-2mm]{0mm}{6mm}
Method of Eq.~\eqref{method3} &
  0.x47642\,74083\,27179(5) &
  0.x44252\,00451\,24(7) \\
\rule[-2mm]{0mm}{6mm}
Continued fraction &
  0.x47642\,74083\,271(9) & 
  0.x44252\,00451\,24688\,3662(3)\\
\hline
\hline
\rule[-2mm]{0mm}{6mm}
$-g$ & 
\multicolumn{1}{c}{$5.0$} &
\multicolumn{1}{c}{$21.6$} \\
\hline
\rule[-2mm]{0mm}{6mm}
Method of Eq.~\eqref{method1} &
  0.x43389\,06678\,10363\,12813\,1(1) &
  0.x55405\,35184\,61013\,80317\,898(0)  \\
\rule[-2mm]{0mm}{6mm}
Method of Eq.~\eqref{method2} &
  0.x43389\,06678(1) &
  0.x55405\,3518(5) \\
\rule[-2mm]{0mm}{6mm}
Method of Eq.~\eqref{method3} &
  0.x43389\,0667(9) & 
  0.x55405\,35(2) \\
\rule[-2mm]{0mm}{6mm}
Continued fraction &
  0.x43389\,06678\,10363\,12813\,116(9) &
  0.x55405\,35184\,61013\,80317\,898(0) \\
\hline
\hline
\end{tabular}
\end{center}
\end{minipage}
\end{center}
\end{table*}

\begin{table*}
\begin{center}
\begin{minipage}{17cm}
\begin{center}
\caption{\label{table4}
The imaginary part $\Im\, E(g)$ at transformation order $150$
shows the same characteristics as the real part discussed in 
Table~\ref{table3}.} 
\begin{tabular}{c@{\hspace{0.5cm}}gg}%{c..}
\hline
\hline
\rule[-2mm]{0mm}{6mm}
$-g$ & 
\multicolumn{1}{c}{$0.5$} & 
\multicolumn{1}{c}{$1.0$} \\
\hline
\rule[-2mm]{0mm}{6mm}
Method of Eq.~\eqref{method1} &
  0.x00026\,6662(1)& 
  0.x01551\,79258\,2059(4)   \\
\rule[-2mm]{0mm}{6mm}
Method of Eq.~\eqref{method2} &
  0.x00026\,66618\,8(2) & 
  0.x01551\,79258\,2(0) \\
\rule[-2mm]{0mm}{6mm}
Method of Eq.~\eqref{method3} &
  0.x00026\,66618\,82408(1) & 
  0.x01551\,79258\,20(6) \\
\rule[-2mm]{0mm}{6mm}
Continued fraction &
  0.x00026\,66618\,824(6)  & 
  0.x01551\,79258\,20594\,2572(2)\\
\hline
\hline
\rule[-2mm]{0mm}{6mm}
$-g$ & 
\multicolumn{1}{c}{$5.0$} & 
\multicolumn{1}{c}{$21.6$} \\
\hline
\rule[-2mm]{0mm}{6mm}
Method of Eq.~\eqref{method1} &
  0.x18385\,80861\,86171\,172893(3) &
  0.x35140\,17775\,93691\,93624\,451(6) \\
\rule[-2mm]{0mm}{6mm}
Method of Eq.~\eqref{method2} &
  0.x18385\,80861(9) & 
  0.x35140\,1777(6) \\
\rule[-2mm]{0mm}{6mm}
Method of Eq.~\eqref{method3} &
  0.x18385\,8086(0) & 
  0.x35140\,18(0) \\
\rule[-2mm]{0mm}{6mm}
Continued fraction &
  0.x18385\,80861\,86171\,17289\,33(1) &
  0.x35140\,17775\,93691\,93624\,45(1) \\
\hline
\hline
\end{tabular}
\end{center}
\end{minipage}
\end{center}
\end{table*}

%
% Negative real axis
%
\subsection{Negative real axis}

Results for four values of $g<0$ are displayed in Tables~\ref{table3} 
and~\ref{table4}. From the numerical evidence, we conclude that ODM methods
with  $\alpha = 5/2$ converge also for $g<0$ but, from the theoretical analysis,
the convergence is expected to be poorer than for $g>0$, in agreement with
the data. The ODM method~\eqref{method2} 
with $\alpha = 5/4$ converges also on the real negative
axis for $|g|$ large enough, that is, $|g|>0.53$ approximately.
 
From the analysis of Ref.~\cite{rJMPZJU}, we know that the imaginary part
itself is an analytic function with singularities on the real negative axis
only at $g=0$ and at infinity. Moreover, it is a simple positive decreasing
function.  Compared with the method with $\alpha=5/2$, the positivity of the
imaginary part for $g=-|g|+\ii 0$ can again be verified, but  with higher
precision for $g$ large. For $|g|\to\infty$ and $\Im (g) = 0_+$,  from the leading
term in Eq.~\eqref{eODMvexpand}
  suitably rotated into the complex plane, one now infers 
\begin{align}
& E(g)/|g|^{1/5} \mathop{=}_{g\to-\infty}
0.3013958756586835717823(7)  
\nonumber\\[2ex]
& \qquad \; +0.2189769214314493762936(0) \; \ii\,, 
\end{align}
a result that uses the perturbative series up order 150.

For $g=-0.5$, the method~\eqref{method2}
with $\alpha=5/4$ is not expected to converge and,
indeed, numerical indications are that this is the case. The result is somewhat
equivalent to a direct summation of the initial asymptotic series, with an
error of the order of the imaginary part on the cut. However, compared to the
direct summation of the series, in some range the successive results first
oscillate around the exact value with the order $K$ with a rather slowing
increasing amplitude rather than blowing up. 
Otherwise, we notice a situation comparable to the real
positive axis. For $|g|< 1$, the ODM method of~\cite{rJMPZJU} yields the
most precise results, while for $|g|\ge 1$, the 
method~\eqref{method1} with $\alpha=5/4$ and convergence acceleration takes
over. However, the continued fraction of Sec.~\ref{ssODMcontfrac} gives the most precise results in the intermediate range and is equivalent for $|g|$ large. 

%
% Strong coupling and analytic properties 
%
\section{STRONG COUPLING EXPANSION AND LEVEL CROSSING}
\label{strongcoupling}

We here demonstrate that considerable information 
on subleading corrections to the strong coupling asymptotics can be 
obtained by investigating the ODM resummed weak-coupling expansion
of the energy levels.
  
In order to explore more thoroughly the strong coupling expansion and the
analytic properties of the ground state energy of the cubic Hamiltonian, it is very useful to also
consider the  linear  coupling Hamiltonian~\eqref{eODMxiiixi},
\begin{equation}
H^\qqc = -\frac12 \, \frac{\partial^2}{\partial x^2} + 
\ii \, \left( \frac16 \, x^3+ \frac12 \, \chi \; x  \right) \, , 
\end{equation}
because a perturbative expansion in the variable
\begin{equation}
\chi = g^{-4/5} \,,
\end{equation}
of the  energy level of the Hamiltonian $H^\qqc$,  defined in
Eq.~\eqref{eODMxiiixi}, is equivalent to a  large $g$ expansion of the
corresponding  energy of the Hamiltonian~\eqref{eCubic}.  We recall here that
the two Hamiltonians are only equivalent for $\chi>0$, the negative $\chi$ axis
corresponding to $\arg (g) = 5\pi /4$.

For real $g$, the ground state energy  $E(g)$ of the Hamiltonian~\eqref{eCubic}
is a real analytic function. From the numerical evidence provided by the ODM
summation in Ref.~\cite{rJMPZJU}, we conjecture that, that, in the variable
$g^{-1/5}$, the eigenvalues can be continued up to
$|\arg\left(g^{-1/5}\right)|=\pi/4$.  This implies that in the variable $\chi$,
the eigenvalues  $E^\qqc(\chi)$ are real analytic with singularities only on
the negative real axis. The  series in powers of $\chi$ are convergent in a
disk. In the case of the ground state, at the singularity nearest to the origin
(numerical results indicate $\chi=\chi_c= -1.3510\ldots$).  Note that, by
contrast with a Hermitian Hamiltonian, \PT symmetric Hamiltonians do not
experience eigenvalue repulsion, which explains why it is possible for
eigenvalues to merge for $\chi$ real.

\subsection{Strong coupling expansion}
 
We now concentrate on the behaviour of $E(g)$ for $g\to\infty$, or,
equivalently, on the small $\chi$ expansion 
given in Eq.~\eqref{exiiixismallv} of
$E^\qqc(\chi)$:
\begin{equation}
E^\qqc(\chi) = \sum_K E^\qqc_{K} \; \chi^K\,.
\end{equation}
First, we have calculated the values of the coefficients of the large coupling expansion
as determined by the ODM method~\eqref{method1} with $\alpha=5/4$, with the improvement by the algorithm \eqref{edelta}. 
The successive coefficients of the small $\chi$ expansion are related to 
$\phi(\lambda,\rho)$ and its derivatives taken at $\lambda=1$ 
(see Appendix~\ref{paramodm} for the
first terms). At leading order, one finds
\begin{equation}
\phi(1,\rho) = \rho^{1/5} \; E^\qqc_{0}\,.
\end{equation}
For the leading term, we obtain 
$$E^\qqc_{0}=0.37254\,57904\,52207\,09825\,0601(1).$$
Note that even at order 150, with the ODM method~\eqref{method3} for the
coefficient $E^\qqc_{0}$ one still finds an 
uncertainty of $4.0 \times 10^{-6}$. With the
method~\eqref{method2}, one obtains $E^\qqc_{0}= 0.372\,545\,78(9)$.

From the ODM method~\eqref{method1} with $\alpha=5/4$ applied to the
perturbative expansion of the function \eqref{eODMFxiii}, at order 150, we have
then determined, using the algorithm described in Appendix~\ref{determination},
with decreasing relative precision (about $10^{-6}$ for the last term), all
terms up to order $K=20$.  
The analysis of the behaviour of the coefficients with increasing order
strongly suggests the existence of a square root singularity located at
$\chi=\chi_c=-1.351\pm0.002$. A square root singularity is consistent with the
existence of a level merging. To confirm this analytic structure and locate
the singularity more precisely, we have assumed that the point $\chi_c$
corresponds to a merging between the ground state energy   and the first
excited state energy.
%

%%%%%%%%%%%%%%%%%%%%%%%%
\subsection{Level merging and strong coupling expansion}

In addition to the ground state, as described, 
we have also generated the perturbative series for the first excited state. We
do not report here the details of the numerical study of the energy of the
first excited state. The behaviour of the series with respect to the ODM summation
methods is very similar. The coefficients of the strong coupling expansion can
be determined with comparable precision. The convergence properties of the ODM
method \eqref{method3} indicate that, again, the eigenvalue is analytic in a
cut-plane and that the first singularity corresponds to the level merging with
the ground state energy. 

We have then formed the two symmetric combinations, the half sum $S_{01}$ and
the half difference squared $\Delta_{01}$ ($\textrm{GS}$ and $\textrm{ES}$
stand for ground state and first excited state, respectively): 
\begin{equation}
\Delta_{01}=\textstyle{\frac{1}{4}}
\left(E^\qqc_{\textrm{GS}}-E^\qqc_{\textrm{ES}}\right)^2\,,
\quad S_{01}=\textstyle{\frac{1}{2}}
\left(E^\qqc_{\textrm{GS}}+E^\qqc_{\textrm{ES}}\right).
\end{equation}
If the ansatz is correct, the two functions are not singular  at $\chi_c$ and
$\Delta_{01}$ must vanish linearly at $\chi_c$. This is indeed what is observed
(see Figs.~\ref{fig6} and~\ref{fig7}).  The result confirms that at
$\chi_c$ the eigenvalues corresponding to the ground state and the first
excited state, merge and for $\chi<\chi_c$ become complex conjugate.

More precisely, the direct summation of the series for $\Delta_{01} $ with the
ODM method of Sec.~\ref{optimizedmap}  shows that $\Delta_{01}$ vanishes
linearly with $\chi$ at the point 
\begin{equation}
\label{chic}
\chi_c=-1.3510415966(3) \,,
\end{equation}
a result fully
consistent with the direct study of the ground state. At this point,
$E^\qqc(\chi_c)=0.41330579447(3) $. As a by-product, one also obtains for the
ground state, the value of  $E^\qqc (\chi=-2^{4/5})=0.38985 020(5)- 0.36442
7(9)\ii $, which confirms the result coming from the continued fraction in
Table \ref{table6}.

By the same method as for the ground state, we have determined the strong
coupling expansions of $S_{01}$ and $\Delta_{01}$. From these strong coupling
expansions, one can recalculate the strong coupling expansion  of the ground
state energy. The results are completely consistent with those from the direct
expansion. However, the precision is improved for the higher order coefficients
as expected since the singularity at $\chi_c$ is now explicitly generated.

As a necessary ingredient for a more precise determination of the 
  coefficients of the continued fraction of Sec.~\ref{ssODMcontfrac} for the ground state,
we have calculated the strong-coupling expansions of the difference $\Delta_{01}$ and the sum $S_{01}$ up to to order 28. We have then inferred the strong coupling expansion of the ground state at the same order although we here give 
only the first 20  terms. These are useful for reference
purposes and read as follows: 
\begin{align}
\label{eODMvexpand}
& E^\qqc(\chi) =
0.37254\,57904\,52207\,09825\,06011(5)
\nonumber\\
& + 0.36753\,58055\,44193\,60353\,04(6)\; \chi 
\nonumber\\
& +0.14378\,77004\,15066\,51583\,39(0)\; \chi^2 
\nonumber\\
& -0.02658\,61056\,27059\,38713\,52(9)  \; \chi^3 
\nonumber\\
& + 0.00988\,71650\,79200\,88729\,05(5)\; \chi^4 
\nonumber\\
& -0.00461\,00192\,93623\,15160\,2(3) \; \chi^5 
\nonumber\\
& + 0.00240\,93426\,35048\,47521\,1(7) \; \chi^6 
\nonumber\\
& -0.00134\,88515\,29319\,85498(8)  \; \chi^7 
\nonumber\\
& + 0.00079\,06119\,76816\,97837(2)  \; \chi^8 
\nonumber\\
& -0.00047\,88478\,41414\,5725(4)  \; \chi^9 
\nonumber\\
& +0.00029\,72375\,58426\,7145(5) \; \chi^{10} 
\nonumber\\
& -0.00018\,80657\,95326\,713(9) \; \chi^{11} 
\nonumber\\
& +0.00012\,08255\,49560\,587(6) \; \chi^{12} 
\nonumber\\
& -7.86045\,58627\,946(5)  \times 10^{-5} \; \chi^{13} 
\nonumber\\
& +  5.16744\,64642\,199(1) \times 10^{-5} \; \chi^{14}
 \nonumber\\
& - 3.42729\,47828\,030(3) \times 10^{-5} \; \chi^{15} 
\nonumber\\
& + 2.29050\,20869\,87(5) \times 10^{-5} \; \chi^{16} 
\nonumber\\
&-1.54091\,59219\,76(3)\times 10^{-5} \; \chi^{17} 
\nonumber\\
& +  1.04266\,03452\,042(4) \times 10^{-5} \; \chi^{18} 
 \nonumber\\
& -0.70913\,87535\,56(4)\times 10^{-5} \; \chi^{19} 
\nonumber\\
& + 0.48450\,81660\,66(0)\times 10^{-5} \; \chi^{20} 
+ \calO(\chi^{21}) \,.
\end{align}
Note that the errors are strongly correlated.

This expansion is also consistent with the  20 first coefficients  reported in
Ref.~\cite{rZnojil}, which have been determined  with a $10^{-10}$--$10^{-11}$
relative precision by a numerical solution of the eigenvalue equation. It is
also consistent with results obtained for the few first terms from a numerical
solution of the Schr\"{o}dinger equation~\cite{rUJJZJxiii}, which have
relative errors of order $10^{-9}$.

\begin{table*}
\caption{\label{table5}
Coefficients $a_p$ of the continued fraction for $K=150$. The singularity is consistent with a limit $0.1850424\ldots$. } 
\begin{tabular}{hhhh}
\hline
\hline
\multicolumn{1}{l}{
\rule[-2mm]{0mm}{6mm}
$p = 1$} & 
\multicolumn{1}{l}{$p = 2$} & 
\multicolumn{1}{l}{$p = 3$} & 
\multicolumn{1}{l}{$p = 4$} \\
\rule[-2mm]{0mm}{6mm}
0.x39122\,09320\,72635\,98993  & 
 0.x18489\,83296\,22869\,56168  &
 0.x18699\,38616\,55333\,76095  & 
 0.x18768\,40668\,91881\,09149 \\
\hline
\multicolumn{1}{l}{
\rule[-2mm]{0mm}{6mm}
$p = 5$} &
\multicolumn{1}{l}{$p = 6$} &
\multicolumn{1}{l}{$p = 7$} &
\multicolumn{1}{l}{$p = 8$} \\
\rule[-2mm]{0mm}{6mm}
0.x18519\,10686\,95077\,9010(9)  & 0.x18477\,49761\,41944\,774(1)  &
0.x18470\,22248\,17836\,84(5)   & 0.x18507\,38286\,03338\,2(9)  \\
\hline
\multicolumn{1}{l}{
\rule[-2mm]{0mm}{6mm}
$p = 9$} &
\multicolumn{1}{l}{$p = 10$} &
\multicolumn{1}{l}{$p = 11$} &
\multicolumn{1}{l}{$p = 12$} \\
\rule[-2mm]{0mm}{6mm}
0.x18513\,42403\,94894(8)  & 0.x18510\,36391\,9575(7)  & 0.x18499\,98937\,209(1)    & 0.x18501\,58945\,34(4) \\
\hline
\multicolumn{1}{l}{
\rule[-2mm]{0mm}{6mm}
$p = 13$} &
\multicolumn{1}{l}{$p = 14$} &
\multicolumn{1}{l}{$p = 15$} &
\multicolumn{1}{l}{$p = 16$} \\
\rule[-2mm]{0mm}{6mm}
0.x18504\,17233\,0(4)  & 0.x18506\,43232\,9(6) & 0.x0.18504\,19101(5)   & 0.x18503\,66878(1) \\
\hline
\hline
\end{tabular}
\end{table*}

% Continued fraction
\section{Continued fraction and analytic properties}
\label{analytic}

%
% General remarks
%
\subsection{General remarks}

To discuss more thoroughly the analytic properties of the ground state energy,
it is convenient to consider the  Hamiltonian \eqref{eODMxiiixi} and
parameterize the energy in terms of the coupling $\chi$.

Inverting the relations~\eqref{exiiilargeg} and \eqref{exiiixismallv}, one
finds a form relevant for the large $\chi$ behaviour of 
$E^\qqc(\chi)$:
\begin{equation}
\label{eODMepslarge} 
E^\qqc(\chi) = \frac{1}{3}\chi^{3/2}+\chi^{1/4}E(\chi^{-5/4}) \, .
\end{equation}
The perturbative expansion  of $E(g)$ in $g$ translates into a large $\chi$
expansion, because large $\chi$ corresponds to weak coupling $g$,
\begin{align}
\chi^{1/4} E (\chi^{-5/4}) =& \;
\frac{1}{2} \, \chi^{1/4} + \frac{11}{288} \, \chi^{-1}
\nonumber\\
& \; -\frac{155}{13824} \, \chi^{-9/4} + \cdots\, \cdot
\end{align}
In particular, the imaginary part for $\chi=-|\chi|+\ii \, 0$ has the expansion
\begin{align}
\Im \; E^\qqc(\chi) =& \;
-\frac{1}{3}|\chi|^{3/2}+\frac{1}{2\sqrt{2}}|\chi|^{1/4}
\nonumber\\[2ex]
& \; + \frac{155}{13824\sqrt{2}}|\chi|^{-9/4} + \cdots \,\cdot
\end{align}
For $|\chi|\gg 1$, the imaginary part is clearly negative. The question that
arises is its sign on the cut for $|\chi|$ of order unity. 

Note that the large $\chi$ behaviour \eqref{eODMepslarge} implies that the Cauchy
representation for the perturbative expansion coefficient $E^\qqc_K$
can be written as
\begin{equation}
E^\qqc_K =
\frac{1}{\pi} \int_{-\infty}^{\chi_c}
\frac{\Im \; E^\qqc(\chi)}{\chi^{K+1}} \, \dd \chi\,,
\end{equation}
converges only for $K > 1$.
The conjecture $\Im \; E^\qqc(\chi) \le 0$, which will be substantiated in Sec.~\ref{ssODMcontfrac},  is thus consistent with the signs of
coefficients of the expansion~\eqref{eODMvexpand}.
Finally, some additional information can be obtained directly from the ODM
method with $\alpha=5/2$ (see Table~\ref{table6}) and also by summing the
small $\chi$ expansion \eqref{eODMvexpand} as we will show.

\subsection{Continued fraction}
\label{ssODMcontfrac}

To continue the expansion~\eqref{eODMvexpand} outside the circle of
convergence, we introduce the continued fraction expansion
of
\begin{equation}
\tilde E^\qqc(\chi) =
\frac{E^\qqc(\chi)- E^\qqc(0)}{ {E^\qqc}'(0) \chi}
\end{equation}
motivated by the prejudice that $\tilde E^\qqc$ is a Stieltjes function since
\begin{equation}
\Im \; \tilde E^\qqc(\chi) = 
\frac{\Im E^\qqc(\chi)}{{E^\qqc}'(0) \; \chi}
\end{equation}
is positive if $\Im E^\qqc_0$ is negative and $\tilde E^\qqc$ behaves like
$\chi^{1/2}$ for $|\chi|\to \infty$.
We then construct  the continued fraction expansion for
the function $\tilde E^\qqc$ by introducing the recursion relation 
\begin{equation}\label{econtfracglarge}
f_{p-1}(\chi) = 1 + \frac{a_p \; \chi }{ f_p(\chi)} \,,\quad f_p(0) = 1 \,,
\end{equation}
with the initial condition $f_0(\chi) = \tilde E^\qqc(\chi)$. It allows us to
calculate the coefficients $a_p$ recursively. The truncated continued
fractions, obtained by replacing at some order $f_p(\chi)$ by 1, generate,
alternatively, $[n+1 / n]$ and $[n / n]$ Pad\'e approximants, depending on
whether $p$ is odd or even.

For a Stieltjes function, all coefficients $a_p$ are positive. Indeed,  using
the coefficients \eqref{eODMvexpand}, which are obtained from the series at
order 150, we find $a_p > 0$ (see Table~\ref{table5}) for $p\le27$  and   the
coefficients seem to converge slowly, within errors, toward the limit expected
for a square root singularity at $\chi=\chi_c$:
$1/(4\chi_c)=-0.18504241514\ldots$. For example, the two last coefficients
determined with sufficient precision are $a_{24}=0.1850464(8)$ and
$a_{25}=0.185042(8)$.

To test the idea further, we have substituted in the
infinite continued fraction 
\begin{equation}
\label{level}
f_{p_{\max}}(\chi)\mapsto \frac{1}{2}+\frac{1}{2}\sqrt{1+4 \gamma \chi} \,,
\end{equation}
$p_{\max}$ varying from $9$ to 27.  
This amounts to taking $a_p$ constant, $a_p=\gamma\approx0.1850$ for $p>p_{\max}$. 
Remarkably enough, for $\chi\to\infty$, the corresponding approximation for
$E^\qqc(\chi)$ behaves like $\chi^{3/2}$ with a coefficient close to the exact
value $1/3$.  We have thus adjusted $\gamma$ more precisely to get $1/3$
exactly. This yields a very stable sequence up to $p=25$, converging slowly toward the expected limit $0.18504\ldots$:
\begin{align}
\gamma_{17} =& \; 0.1850687 \,, \  
\gamma_{18} = 0.1850261 \,, \ 
\gamma_{19} = 0.1850603  \,,
\nonumber\\[2ex]
\gamma_{20} =& \; 0.1850239 \,, \ 
\gamma_{21} = 0.1850562 \,, \ 
\gamma_{22} = 0.1850310 \,,
\nonumber\\[2ex]
\gamma_{23} =& \; 0.1850562 \,,\ \gamma_{24}=0.1850367\,,\ \gamma_{25}=0.185048(4).
\end{align}
corresponding to a singularity located at $\chi=\chi_c\approx -1.3504 $,
consistent with a direct analysis of the behaviour of the coefficients in the
expansion \eqref{eODMvexpand}.
As a consequence, in all Tables and Figures we have reported results obtained from
this approximated continued fraction.
 
%
% Spectrum for $\maybebm{\chi<0}$
%
\subsection{Spectrum for $\maybebm{\chi<0}$}

\begin{figure}
\includegraphics[width=0.7\linewidth]{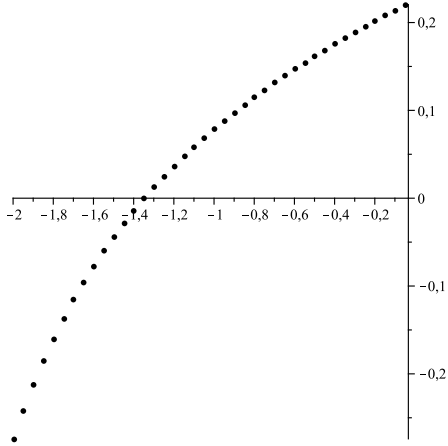}
\caption{\label{fig6} $ \Delta_{01}=(E^{\qqc}_{\textrm{ GS}}-E^{\qqc}_{\textrm{
ES}})^2/4$ as a function of $\chi$ for $-1.85\le \chi\le -0.85$
[the value at the origin is $\Delta_{01}(\chi=0)=0.2263073\ldots$].  } 
\end{figure}

\begin{figure}
\includegraphics[width=0.7\linewidth]{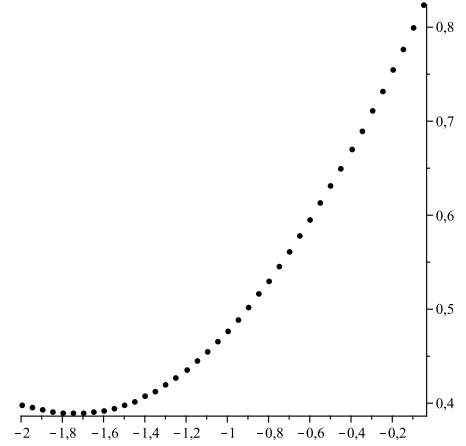}
\caption{\label{fig7} $ S_{01}=(E^{\qqc}_{\textrm{ GS}}+E^{\qqc}_{\textrm{
ES}})/2$ as a function of $\chi$ for $-1.85\le \chi\le -0.85$
[the value at the origin amount to $S_{01}(\chi=0)=0.8482634\ldots$].  } 
\end{figure}

The convergence of the various methods is rather poor for $\chi<0$, specially
in the neighbourhood of the cut. Table~\ref{table6} displays a few results for
$E^\qqc(\chi)$.  According to Table~\ref{table6}, the most precise results are
in general obtained from the modified continued fraction, with a precision that
for $|g|>1$ is comparable with the ODM method defined by  Eq.~\eqref{method1}.
Note that for the first value, which corresponds to $|g|=0.5$, the ODM method
with $\alpha=5/4$ does not converge, as expected, and does not even provide a
reliable estimate. The method of Sec.~\ref{ssODMnewmap} is not useful either.
Finally, in the neighbourhood of the singularity $\chi=\chi_c$ a calculation
based on the determination of $\Delta_{01}$ and $S_{01}$ gives the most precise
results (see  Figs.~\ref{fig6} and \ref{fig7}). 

The imaginary part after the cut for $\chi=-|\chi|+\ii \, 0$ is necessarily
negative when calculated from the approximated continued fraction. Therefore,
the general consistency of the results coming from the continued fraction and
the other ODM methods provides an additional confirmation of our conjecture
about the sign of the discontinuity.

\begin{table*}
\begin{center}
\begin{minipage}{13.5cm}
\begin{center}
\caption{\label{table6}
Values for $E^\qqc(\chi)$, obtained by various 
ODM summation methods, on the real negative axis, at transformation 
order 150. Method~\eqref{method1} and the  continued 
fraction approximation  \eqref{econtfracglarge} provide the best numerical convergence.} 
\begin{tabular}{c@{\hspace{0.3cm}}gg}
% first set of values
\hline
\hline
\rule[-2mm]{0mm}{6mm}
$\chi$ & 
\multicolumn{1}{l}{$-2^{4/5} = -1.741\ldots$} & 
\multicolumn{1}{l}{$-1$} \\
\hline
\rule[-2mm]{0mm}{6mm}
Method of Eq.~\eqref{method1}  &  
\multicolumn{1}{c}{no convergence} & 0.x19575\,08157(1)  \\
\rule[-2mm]{0mm}{6mm}
Method of Eq.~\eqref{method2}  &  
  0.x42 \pm 0.20 - (0.39\pm0.17) \; \ii  & 0.x195(8) \\
\rule[-2mm]{0mm}{6mm}
Method of Eq.~\eqref{method3}  &  
  0.x38(1) - 0.36(8) \; \ii & 0.x195(5) \\
\rule[-2mm]{0mm}{6mm}
Continued fraction 
  & 0.x3898(5)-0.3644(3) \; \ii  & 0.x19575\,08157\,16171\,9(6) \\
%
% second set of values
%
\hline
\hline
\rule[-2mm]{0mm}{6mm}
$\chi$ &
\multicolumn{1}{l}{$-5^{4/5} = -0.275\ldots$} &
\multicolumn{1}{l}{$-( 21.6)^{-4/5}=-0.085\ldots$} \\
\hline
\rule[-2mm]{0mm}{6mm}
Method of Eq.~\eqref{method1}  &  
  0.x28269\,92581\,93274\,90989\,9(0) & 
  0.x34215\,80186\,19340\,42140\,767(3) \\
\rule[-2mm]{0mm}{6mm}
Method of Eq.~\eqref{method2}  &  
  0.x28269\,93(0) & 0.x34215\,80(2) \\
\rule[-2mm]{0mm}{6mm}
Method of Eq.~\eqref{method3}  &  
  0.x2827(1) & 0.x34215(6) \\
\rule[-2mm]{0mm}{6mm}
Continued fraction 
  & 0.x28269\,92581\,93274\,90989\,90(1) 
  & 0.x34215\,80186\,19340\,42140\,767(6)   \\
\hline
\hline
\end{tabular}
\end{center}
\end{minipage}
\end{center}
\end{table*}

% Conclusions
%
\section{Conclusions}
\label{conclu}

Let us briefly summarize the findings of the current numerical study of
properties of the cubic anharmonic oscillator.  For $g \ge 0$, the
Hamiltonian~\eqref{eCubic} is \PT symmetric and has a real positive spectrum.
The energy eigenvalues have divergent, Borel summable expansions in powers of
the coupling constant $g$. The imaginary parts of the eigenvalues on the cut on
the real negative axis are positive. Starting from the \PT symmetric case and
developing a strong coupling expansion in the variable $\sqrt{\xi} =g^{-1/5}$,
one sees that eigenvalues are  univalued functions in some neighbourhood of the
origin in the $\xi$ plane, with a simple pole at $z = \sqrt{\xi} =g^{-1/5}=0$
as a unique singularity.

In Sec.~\ref{summation}, we have discussed various possible summation methods
for the determination of energy eigenvalues of the cubic potential. 
Notably, we explore the numerical properties of three such 
methods here, given in Eqs.~\eqref{method1},~\eqref{method2} 
and~\eqref{method3}. The latter was used in our 
previous paper~\cite{rJMPZJU}.
Summing the perturbative expansion of the ground state energy $E(g)$ by various
implementations of the ODM method, we uncover additional properties. 
From the apparent convergence of the ODM method for $\alpha=5/2$
[see Eq.~\eqref{method3}],
we conclude that in the variable $z$ the function
$E(g(z))$ has no other singularity in the sector $|{\rm arg}\, z|<\pi/4$. In
the variable $\chi=g^{-4/5}=z^4$, this translates into the property that the
ground state energy $E^\qqc(\chi)$ of the Hamiltonian \eqref{eHamxiiixi}
also is a real analytic function in a cut plane with  a cut along the real negative axis.

The ODM summation method~\eqref{method1} converges very well in the strong
coupling regime and thus allows us to determine a number of terms of the large
$g$, thus small $\chi$ expansion. The precision can be further improved by the
convergence acceleration algorithm~\eqref{edelta} and, finally, by a combined
calculation of the ground state and first excited energies.  The first 20 terms
are given in Eq.~\eqref{eODMvexpand}. A direct analysis of the strong coupling
expansion indicates the existence of a square root singularity located at
$\chi=\chi_c\approx-1.351$ [$g\approx 0.687\, \exp(5 \ii\pi/4)$ in the second
sheet]. These properties are confirmed by the direct calculation of the
difference $\left(E_{\textrm{GS}}-E_{\textrm{ES}}\right)^2$, which vanishes
linearly at the point $\chi_c=-1.3510415966(3)$, as given in Eq.~\eqref{chic}. 

The strong coupling expansion, in turn, can be summed by an expansion in a
continued fraction (Sec.~\ref{ssODMcontfrac}). The calculated coefficients of
the continued fraction display an asymptotic behavior
consistent with a square
root singularity. The extrapolated  continued fraction yields results in
remarkable agreement with the more direct ODM calculations.

As a function of $z$, $E^\qqc(\chi(z))$ is invariant 
under the rotation $z\mapsto
z\exp(\ii\pi/2)$. It is thus entirely defined by its values for $-\pi/4\le {\rm
arg}(z) \le +\pi/4$, that is, $-5\pi/4\le {\rm arg}(g) \le +5\pi/4$. As a
consequence, as a function of $z$, $E$ has cuts only on the lines ${\rm arg}\,
z=n\pi/4$ where level merging can occur.  Note that, from the point of view of
Eq.~\eqref{eODMxiiixi}, for $\chi<0$ the Hamiltonian is still \PT symmetric
and, thus, the singularity at $\chi=\chi_c$ corresponds to a kind of
spontaneous \PT symmetry breaking. 

All coefficients of the associated expansion in a continued fraction are found
to be positive as long as we can estimate them with enough precision, that is,
up to order 27 (corresponding to a strong coupling expansion up to order 28), 
and seems to converge toward a positive value consistent with
the square root singularity of the first level merging. This gives a very
strong argument that the eigenvalue of the strong coupling Hamiltonian $\Im
E^\qqc(-|\chi|+ \ii \, 0)$ is negative on the cut. This conjecture is also
consistent with the large $\chi$ behaviour, the Cauchy representation, the
coefficient of the square root singularity. However, this property cannot be
shared by all eigenvalues and may even be unique to the ground state.

To go beyond this study, one would have to apply these ODM summation methods
more directly to the spectral equation, in order to eliminate all level
merging singularities (for the pure cubic potential see Ref.~\cite{rAVoros}).
Finally, we would like to emphasize the empirical evidence
gathered during the current study, for the robustness of the ODM summation
methods~\eqref{method1},~\eqref{method2} and~\eqref{method3}, 
which applied to different functions with varying implementation 
always give consistent results.

%
% Acknowledgments
%
\section*{Acknowledgements}

J.Z.-J.~gratefully acknowledges CERN's hospitality, where a major part of this work was
completed.  U.D.J.~acknowledges support by a Grant from the Missouri Research
Board and by the National Science Foundation (Grant PHY--8555454). 

\appendix{}

% 
% ODM and strong coupling expansion
% 
\section{ODM and strong coupling expansion}
\label{paramodm}

We assume that we know a perturbative expansion for $E(g)$
in powers of the coupling constant $g$, and that 
$E(g)$ also has a strong-coupling asymptotic
expansion of the form,
\begin{equation}
\label{eElargegexpansion}
E(g) = g^\beta \sum_{n=0} \epsilon_n g^{-n/\alpha} \,, \qquad 
g \to \infty\,,
\end{equation}
a property shared by the example~\eqref{eODMxiiixi} we discuss here.
It is also shared 
by the quartic anharmonic oscillator~\cite{rsezzin} and all $x^N$
perturbations to the quantum harmonic oscillator. 
We then consider the conformal mapping \eqref{eODMalpha}, 
\begin{equation}
\label{eODMalphab} 
g = \rho \, \frac{\lambda}{(1 - \lambda)^{\alpha}} \,.
\end{equation}
This transformation maps
the real positive $g$ axis onto the finite $\lambda$ interval $[0,1]$.
For  $g\to\infty$, $\lambda\to1$ and $g$ has an expansion at $\lambda=1$ of the form
\begin{equation}
g^{-1/\alpha}= \sum_{n=0 } \Lambda_n(1-\lambda)^{n+1}
\end{equation}
with $\Lambda_0=\rho^{-1/\alpha}$. The function 
\begin{equation}
\phi(\lambda) = (1-\lambda)^{\alpha\beta} \; E\bigl(g(\lambda)\bigr)  
\end{equation}
then has a Taylor series expansion at $\lambda=0$,
\begin{equation}
\phi(\lambda) = \sum_{n=0} \phi_n \; \lambda^n \,,  
\end{equation}
as well as at $\lambda=1$,
\begin{equation}
\phi(\lambda) = \sum_{n=0} \varphi_n(1-\lambda)^n \,.
\end{equation}
with $\varphi_0 = \epsilon_0 \; \rho^\beta $, where $\epsilon_0$ is the coefficient
defined in~\eqref{eElargegexpansion}. This last property explains, to a large
extent, the good convergence of the method even for $g\to\infty$.

\vspace*{1cm}

%
% Large $g$ expansion: a few terms
%
\section{Large $\maybebm{g}$ expansion: a few terms}
\label{determination}

In order to 
determine the successive terms of the large $g$, small $\chi$ expansion we do
not have to differentiate numerically the values of $E(g)$ calculated at $g$
large but finite. In the ODM method, the various terms have explicit analytic
forms and we give here the first terms.
First, we set
\begin{align}
E(g) =& \; \phi(\lambda)(1-\lambda)^{-\alpha\beta} 
\nonumber\\[2ex]
=& \; \phi(\lambda) \,
\left( \frac{g}{\rho} \right)^{\beta} \, \lambda^{-\beta}
\equiv \left( \frac{g}{\rho} \right)^\beta \psi(\lambda).
\end{align}
Then we write the relation \eqref{eODMalphab} as
\begin{equation}
\lambda = 1 - \left( \frac{\rho}{g} \right)^{1/\alpha} \, \lambda^{1/\alpha}.       
\end{equation}
Setting $(\rho/g)^{1/\alpha}=z$, we expand  $s=1-\lambda$  in powers of $z$,
so that $s \equiv 1-\lambda=z- \frac{1}{ \alpha} \, z^2+\cdots$.
The function $\psi(\lambda)$ can then be expanded successively in powers of $s$
and $z$:
\begin{align}
\psi(\lambda) & =
\psi(1) - \psi'(1) \, s + 
\tfrac12 \psi''(1) \, s^2 + \cdots \nonumber\\[2ex]
& = \psi(1)-\psi'(1) \, z+
\left\{ \frac12 \, \psi''(1) + \frac{\psi'(1)}{\alpha} \right\} 
\, z^2+\cdots
\nonumber\\[2ex]
& = \rho^\beta \; 
\left( \epsilon_0 
+ \rho^{-1/\alpha}\epsilon_1 \, z
+ \rho^{-2/\alpha}\epsilon_2 \, z^2 +
\cdots \right) \,,
\end{align}
where the $\epsilon_n$ coefficients are given 
in Eq.~\eqref{eElargegexpansion}. The first terms yield
\begin{align}
\epsilon_0 =& \; \psi(1) \; \rho^{-\beta} ,
\\[2ex]
\epsilon_1 =& \; -\psi'(1) \; \rho^{-\beta+1/\alpha}, 
\nonumber\\[2ex]
\epsilon_2 =& \;
\left\{ \frac12 \, \psi''(1)+
\frac{\psi'(1)}{\alpha}
\right\} \; \rho^{-\beta+2/\alpha}, 
\nonumber\\[2ex]
\epsilon_3 =& \;
-\left\{ \frac{3 - \alpha}{2 \alpha^2} \; \psi'(1) 
+ \frac{\psi''(1)}{\alpha} 
+ \frac{\psi'''(1)}{6} \right\}
\rho^{-\beta+3/\alpha} \, .
\nonumber
\end{align}

%
% Numerical evidence for strong coupling
%
\section{Numerical evidence for strong coupling}
\label{num_evidence}

In Ref.~\cite{rJSLZ}, we started from the Hamiltonian
\begin{equation}
H = 
- \frac12 \, \frac{\partial^2}{\partial x^2} 
+ \frac12 \, x^2 + \sqrt{g} \, x^3 \,.
\end{equation}
which entails both a change in the normalization of the 
coupling term and also a change in the complex 
phase of the coupling. The transformation
$x \to g^{-1/10} \, x$
brings the Hamiltonian into the form
\begin{equation}
H = g^{1/5} 
\left(
- \frac12 \, \frac{\partial^2}{\partial x^2} 
+ x^3 
+ \frac{x^2}{2 \, g^{5/2}} \right) \,.
\end{equation}
We now define $g = u^{-5/2}$  and write
\begin{equation}
H = g^{1/5}
\left(
- \frac12 \, \frac{\partial^2}{\partial x^2}
+ x^3
+ u \, \frac{x^2}{2} \right) \,.
\end{equation}
A shift of the variable $x \to x - \tfrac16 \, u$
then results in
\begin{align}
\label{strongH}
H =& \; 
g^{1/5} \left( - \frac12 \, \frac{\partial^2}{\partial x^2}
+ x^3 - \frac{u^2}{12} \, x + \frac{u^3}{180} \right) 
\nonumber\\[2ex]
=& \; g^{1/5} \left( - \frac12 \, \frac{\partial^2}{\partial x^2}
+ x^3 - \frac{1}{12} \, g^{-4/5} \, x + \frac{1}{180} \, g^{-6/5} \right) \,.
\end{align}
If one now writes the strong-coupling expansion
of the $N$th energy level in the form of Eq.~(17) of 
Ref.~\cite{rJSLZ},
\begin{equation}
\label{according}
E_N(g) \, = \, g^{1/5} \,
\sum\limits_{K = 0}^{\infty} L_{N,K} \, g^{-2K/5} \,,
\end{equation}
then the terms [see Eq.~\eqref{strongH}]
\begin{equation}
- \frac12 \, \frac{\partial^2}{\partial x^2}
+ x^3 - \frac{1}{12} \, g^{-4/5} \, x
\end{equation}
generate a convergent series in even powers of $g^{-4/5}$,
and the term
\begin{equation}
\frac{1}{180} \, g^{-6/5}
\end{equation}
in Eq.~\eqref{strongH} is the only term of odd power in 
$g^{-2/5}$ in Eq.~\eqref{according}.
From the point of view of the strong coupling expansion,
this term $L_{N,3} = 1/108$ represents
a shift, independent of the quantum number $N$, of all levels of
the cubic potential. This affords an explanation for the 
observations made in Table~2 of Ref.~\cite{rJSLZ}, where 
on the basis of numerical evidence, it 
was conjectured that the terms of order $K = 1$ according Eq.~\eqref{according}
vanish, and that the numerical value for the coefficient of the term of 
order $K =3$ is $1/108$ uniformly for all levels.


\begin{thebibliography}
\expandafter\ifx\csname url\endcsname\relax
  \def\url#1{\texttt{#1}}\fi
\expandafter\ifx\csname urlprefix\endcsname\relax\def\urlprefix{URL }\fi
\expandafter\ifx\csname href\endcsname\relax
  \def\href#1#2{#2} \def\path#1{#1}\fi

\bibitem{rBeBo}
C. M. Bender and S. Boettcher, \textit{Real spectra in 
non-Hermitian Hamiltonian having \PT symmetry}, Phys. Rev. Lett.
80 (1998) 5243-5246.

\bibitem{rDDB}
P. Dorey, C. Dunning and R. Tateo, 
\textit{Spectral equivalences, Bethe ansatz equations, and reality properties in
\PT-symmetric quantum mechanics,} J. Phys. A 34
L391 (2001); \textit{ibid.} 34  (2001) 5679-5704.

\bibitem{rshin}
L. C. Shin, 
\textit{On the reality of eigenvalues for a class of \PT-symmetric oscillators,} 
Commun. Math.  Phys. 229 (2002) 543-564.

\bibitem{rJMPZJU}
J. Zinn-Justin and U. D. Jentschura, 
\textit{Order-dependent mappings: strong coupling behaviour 
from weak coupling expansions in non-Hermitian theories},
J. Math. Phys. 51 (2010) 072106.

\bibitem{rCGM} 
E. Caliceti, S. Graffi, and M. Maioli, 
\textit{Perturbation theory of odd anharmonic oscillators,} 
Commun. Math. Phys. 75 (1980)  51-66.

\bibitem{rECali} 
E. Caliceti, \textit{Distributional Borel summability of odd anharmonic oscillators}, 
J. Phys. A 33 (2000) 3753-3770.

\bibitem{rLOBLip} 
L. N. Lipatov, 
\textit{Divergence of the perturbation-theory series and pseudoparticles}, 
JETP Lett. 25 (1977) 104-107; 
\textit{Divergence of the perturbation-theory series and the quasi-classical 
theory},  Sov. Phys. JETP 45 (1977) 216-223.

\bibitem{rLOBgen}
E. Br\'ezin, J. C. Le Guillou, J. Zinn-Justin, 
\textit{Perturbation theory at large order. I. The $\varphi^N$ interaction}, 
Phys. Rev. D 15 (1977) 1544-1557.

\bibitem{rZJLOreport}
J.~Zinn-Justin, 
\textit{Perturbation series at large orders in quantum mechanics 
and field theories: application to the problem of resummation},  
Phys.~Rep.~70 (1981) 109--167.

\bibitem{rUJASJZJ}
U. D. Jentschura, A. Surzhykov, J. Zinn-Justin, 
\textit{Generalized nonanalytic expansions, \PT--symmetry and large order
formulas for odd anharmonic oscillators}, SIGMA  5 (2009) 005.
 
\bibitem{rVGMMAM}
V. Grecchi, M. Maioli and A. Martinez, 
\textit{Pad\'e summability of the cubic oscillator}, 
J. Phys. A 42 (2009) 425208.
%\nref\rBFM{R. Brower, M. Furman, and M. Moshe, Phys. Lett.
%B 76, 213 (1978).}

\bibitem{rBeWe}
C. M. Bender, E. J. Weniger, 
\textit{Numerical evidence that the perturbation expansion
for a non-Hermitian \PT-symmetric Hamiltonian is Stieltjes}, 
J. Math. Phys. 42 (2001) 2167-2183.

\bibitem{rsezzin}
R. Seznec, J. Zinn-Justin, 
\textit{Summation of divergent series by order dependent mappings: 
Application to the anharmonic oscillator and critical exponents in field theory}, 
J. Math. Phys. 20 (1979) 1398--1408.

\bibitem{rZJODM}
J.~Zinn-Justin, 
\textit{Summation of divergent series: Order-depen\-dent mapping}, 
arXiv:1001.0675 [math-ph].

\bibitem{rZnojil}
F. M. Fernandez, R. Guardiola, J. Ros and  M. Znojil, \textit{Strong-coupling expansions for the \PT-symmetric oscillators
$V(x)=a(\ii x)+b(\ii x)^2+c(\ii x)^3$}, J. Phys. A: Math. Gen. 31 (1998) 10105-10112.

\bibitem{rUJJZJxiii}
U. D. Jentschura,  J. Zinn-Justin, 
\textit{Calculation of the Characteristic Functions of Anharmonic 
Oscillators}, ArXiv:1001.4313 [math-ph].

\bibitem{rCBTTW}
C.M. Bender and T.T. Wu, \textit{Anharmonic oscillator}, Phys. Rev. 184  (1969) 
1231-1260.

\bibitem{rEDDTT}
E. Delabaere and D. T. Trinh, 
\textit{Spectral analysis of the complex cubic oscillator}, 
J. Phys. A 33 (2000) 8771-8796. 

\bibitem{rHKlWJ} 
H. Kleinert, W. Janke, \textit{Convergence behavior of
variational perturbation expansion: A method for locating Bender-Wu
singularities,} Phys. Lett. A 206 (1995) 283-289.

\bibitem{rLYFisher}
M. E. Fisher, 
\textit{Yang--Lee Edge Singularity and $\phi^3$ Field Theory}, 
Phys. Rev. Lett. 40 (1978) 1610-1613.

\bibitem{rBBJ}
C. M. Bender, D. C. Brody and H.F. Jones, 
\textit{Scalar Quantum Field Theory with Cubic Interaction}, 
arXiv: hep-th/0402011.

\bibitem{rJSLZ}
U. D. Jentschura, A. Surzhykov, M. Lubasch, and J. Zinn-Justin, 
\textit{Structure, Time Propagation and Dissipative Terms for Resonances}, 
J. Phys. A 41 (2008) 095302.

\bibitem{rMaso}
D. Masoero, 
\textit{Poles of Int\'egrale Tritronqu\'e\'ee and Anharmonic Oscillators. 
A WKB Approach}, arXiv:0909.5537v2 [math.CA].

\bibitem{rRGKKHS}
R. Guida, L. Konishi, H. Suzuki, 
\textit{Improved Convergence Proof of the Delta Expansion and 
Order Dependent Mappings}, Ann. Phys. (N.Y.) 249 (1996) 109-145.

\bibitem{rAVoros} 
A. Voros, \textit{Airy function --- exact WKB results for potentials of odd
degree}, J. Phys. A 32 (1999) 1301-1311. 

\end{thebibliography}
\end{document}